\documentclass[12pt,preprint,numberedappendix]{aastex}

\usepackage{color}
\usepackage[normalem]{ulem}  

\newcommand{\midtilde}{\raisebox{-0.25\baselineskip}{\textasciitilde}}

\begin{document}

\title{New Hyperon Equations of State for Supernovae and Neutron Stars in 
Density-dependent Hadron Field Theory}
\author{Sarmistha Banik} 
\affil{BITS Pilani, Hyderabad Campus, Hyderabad-500078, India}
\author{Matthias Hempel} 
\affil{Departement Physik, 
Universit\"at Basel, Klingelbergstrasse 82, 4056 Basel, Switzerland}
\author{Debades Bandyopadhyay\altaffilmark{1}} 
\affil{Astroparticle Physics and Cosmology Division, Saha Institute of Nuclear 
Physics, 1/AF Bidhannagar, Kolkata-700064, India}
\altaffiltext{1}{Also at Centre for Astroparticle Physics, Saha Institute of Nuclear
Physics, 1/AF Bidhannagar, Kolkata-700064, India}

\begin{abstract}
We develop new hyperon equation of state (EoS) tables for core-collapse 
supernova simulations and neutron stars. These EoS tables are based on a 
density-dependent
relativistic hadron field theory where baryon-baryon interaction is 
mediated by mesons, using the parameter set DD2 from \citet{typ10} for nucleons.
Furthermore, light and heavy nuclei along with the interacting nucleons are
treated in the nuclear statistical equilibrium
model of Hempel and Schaffner-Bielich which includes excluded volume
effects. Of all possible hyperons, we consider only the contribution of 
$\Lambda$s. We have developed two variants of hyperonic EoS tables: in the 
np$\Lambda \phi$ case the repulsive hyperon-hyperon interaction mediated by the
strange $\phi$ meson is taken into account, and in the np$\Lambda$ case it is not.
The EoS tables for the two cases encompass wide range of
density ($10^{-12}$ to $\sim$ 1 fm$^{-3}$), temperature (0.1 to 158.48 MeV), and 
proton fraction (0.01 to 0.60). 
The effects of $\Lambda$ hyperons on thermodynamic quantities such as free 
energy per baryon, pressure, or entropy per baryon are 
investigated and found to be significant at higher densities. 

The cold, $\beta$-equilibrated EoS (with the crust included self-consistently) 
results in a 2.1 M$_{\odot}$ maximum mass neutron star for the np$\Lambda \phi$ 
whereas that for the np$\Lambda$ case is 1.95 M$_{\odot}$. 
The np$\Lambda \phi$ EoS represents the first supernova EoS table 
involving hyperons that is directly compatible with the recently measured 
2 M$_{\odot}$ neutron stars. 

\end{abstract}


\section{Introduction}
Compact astrophysical objects are born in the aftermath of massive stars 
($>$ 8 M$_{\odot}$) through  core-collapse supernova (CCSN) explosions 
in the penultimate stage of their evolution \citep{Bet}. In the CCSN mechanism, 
the gravitational collapse of the iron core begins as the core exceeds the
Chandrasekhar mass. The subsequent core bounce occurs when the core density 
reaches beyond normal nuclear matter density and a hydrodynamic shock is 
generated. If the shock wave is strong enough, this might lead to a prompt 
supernova explosion, which, however, is not found in recent state-of-the-art 
computer simulations. 
The hot and neutrino-trapped protoneutron star (PNS)
settles into hydrostatic equilibrium immediately after the core bounce. The 
PNS could evolve either into a neutron star or into a black hole within
a few seconds after the emission of neutrinos. Though the CCSN explosion 
mechanism has been explored for the past five decades, a complete
understanding of this phenomenon is still beyond our reach. In most CCSN
simulations, the shock stalls after traveling a few hundred kilometers. The 
revival of the shock by neutrino heating \citep{wil} or the generation of a 
second shock due to a first order hadron-quark phase transition \citep{irina} 
could trigger a delayed CCSN explosion. Regarding the latter, until now this
mechanism was only shown to be working for equations of state (EoS) 
that are not compatible with
the latest neutron star mass measurements such as those from \citet{anto}.

Besides the dimensionality of the problem \citep{nord} and neutrino reaction rates, 
the EoS of matter plays a tremendous role in a successful CCSN explosion
\citep{janka12}. The
first nuclear EoS table suitable for CCSN simulations
was formulated by \citet{wolf} 
followed by the Lattimer and Swesty (LS) EoS \citep{ls} and the Shen EoS 
\citep{shen}. 
The last two EoS tables describe all possible compositions of matter depending
on wide ranges of density, temperature, and proton fraction such as free
nucleons, light nuclei in coexistence with nucleons, the ideal gas of nuclei,
and uniform nuclear matter. 
The LS EoS table is based on Skyrme interaction 
for uniform matter and a compressible liquid drop model for non-uniform matter. 
On the other hand, for the first time, the Shen EoS table was constructed using
the relativistic field theory for low- and high-density uniform matter. 
Non-uniform matter was described by the Thomas-Fermi model. Both of these 
two approaches, LS and Shen, employed the single nucleus approximation
and neglected shell effects. The LS and Shen EoS tables have been
used extensively for CCSN simulations over the years. 

Recently, several new EoS were developed, keeping in pace with
updated knowledge from nuclear structure, experimental data, 
or neutron star observations, aiming at an improved underlying description 
and with possibly new particle degrees of freedom taken into account
\citep{hs1,raduta10,horo,horo1,horo2,fis1,bli,hs2,stei,fis2,buyu,toga}. 
One such notable nuclear EoS called the HS EoS was formulated
within the framework of the nuclear statistical equilibrium (NSE) model \citep{hs1}. 
The HS EoS table 
treated the ensemble of nuclei and nucleons in the NSE model using 
the relativistic mean field model for interacting nucleons, incorporated 
excluded volume effects in the thermodynamically consistent manner, considered 
excited states of nuclei and matched the low density matter with
uniform matter at high density \citep{hs1}. A new 
nuclear EoS table was generated
adopting the virial expansion for a non-ideal gas of nucleons and nuclei
by \citet{horo}. The statistical model by \citet{botvina04,botvina10,buyu}, 
is based onthe  multifragmentation of nuclei in heavy-ion collisions. 
In \citet{buyu_compare}, it was compared with some of the other aforementioned 
approaches. 
For the first time, the EoS has been constructed in a variational
calculation using bare nuclear forces such as Argonne v18 (AV18) and Urbana IX
(UIX) by \citet{toga,const}, which, however, does not yet include the case of
non-uniform matter.          

The EoS described above would not only influence the supernova
dynamics but also the formation of neutron stars and their structures. Neutron
star observations could provide important inputs in the construction of EoS
tables for CCSN simulations. The first supernova EoS table directly 
based on measured masses and radii of neutron stars was 
developed by Steiner and collaborators \citep{stei}.
Unlike radii, neutron star masses have been estimated to a very high degree of
accuracy. This has been possible because post-Keplerian parameters, such as
orbital decay, periastron advance, Shapiro delay, and time dilation have been 
measured in many pulsars. Currently, the accurately measured highest neutron 
star mass is 2.01$\pm 0.04$ M$_{\odot}$ \citep{anto}. This puts a strong 
constraint on the $\beta$-equilibrated EoS. Most of the nuclear EoS mentioned 
above result in 2 M$_{\odot}$ neutron stars.

Observed neutron star masses are also probes of compositions of dense matter.
It has long been debated whether or not novel phases of matter such as hyperons, 
Bose-Einstein condensates of kaons, and quarks may exist in neutron star 
interior. It may happen that the phase transition from nuclear matter
to exotic matter could occur in the early post-bounce phase of a CCSN. Strange
degrees of freedom would be crucial for the long-term evolution of the PNS. It is 
to be noted that strange matter typically makes the EoS softer resulting in a 
smaller 
maximum mass neutron star than that of the nuclear EoS \citep{glen}. 
\citet{vandalen14} showed that the observed high masses of neutron stars in combination with hypernuclear
data put tight constraints on the interactions of hyperons in neutron star matter.
Note that there is also an interesting interplay between the strangeness 
content and the symmetry energy on properties of neutron stars, which was 
recently discussed by \citet{provi2013} for the case of hyperonic EoS. 

Several EoS including quark and hyperon matter were developed for and
applied to supernova simulations 
\citep{ishi,naka08,irina,sumi,shen11,naka,oertel12,peres,sb}. 
None of the EoS tables with
exotic matter were directly compatible with the 2 M$_{\odot}$ neutron star
or they were just barely acceptable. On the other
hand, many model calculations including exotic matter such as hyperons showed
that the EoS of $\beta$-equilibrated matter may lead to 2 M$_{\odot}$ or more
massive neutron stars \citep{weis1,weis2,last,colu,lopes,gus,vandalen14}. 

Recently, \citet{fis2} published an quark-hadron hybrid EoS with a
maximum mass above 2 M$_\odot$, which, however, did not lead to a 
phase-transition-induced explosion.
The limited number of realistic supernova EoS with exotic
degrees of freedom motivates
us to construct a hyperon EoS in the relativistic mean field theory with 
density-dependent couplings that is compatible with a 2 M$_{\odot}$ mass 
neutron star. 

The paper is organized as follows. Section 2 describes the 
methodology for the calculation of EoS tables including $\Lambda$ hyperons. 
The results of hyperon EoS tables are discussed in Section 3. Section 4 gives 
a summary and conclusions. 
In the Appendix, we give detailed information about the definition of 
the various quantities stored in the final EoS tables and discuss their 
accuracy and consistency.

\section{Methodology}
Here we describe the models to construct the temperature-dependent hyperon EoS 
spanning over
different regimes of baryon number density, temperature, and proton fraction.
Compositions of matter vary from one region to the other. Constituents of
matter are nuclei, (anti-)neutrons, (anti-)protons, 
(anti-)$\Lambda$ hyperons, electrons, positrons,
and photons. 
We make the standard assumption that 
electrons and positrons form a uniform background in this calculation.
We do not include the contributions of muons, because in
standard core-collapse supernova simulations the net muon lepton fraction
is zero. If desired, muons could be added to the EoS as another 
non-interacting particle species.
The contribution of the neutrinos is similarly not
taken into account in the EoS. These are typically handled by neutrino transport, because
weak equilibrium is generally not obtained. 
In the following, we discuss various models to compute the EoS of matter 
in different regimes, 
where we restrict the discussion on the non-trivial baryonic contribution.

\subsection{Density-dependent Relativistic Mean-Field Theory for Baryons}
\label{rmf}
The relativistic mean field (RMF) model with density-dependent couplings
is adopted for interacting baryons in this calculation. 
We exploit this density-dependent RMF model for a transition from 
non-uniform nuclear to $\Lambda$ hyperon matter. 
The baryon-baryon interaction in this model is
mediated by the exchange of $\sigma$, $\omega$, and $\rho$ mesons. 
The model may also be extended to include hyperon-hyperon interaction 
through hidden-strangeness mesons---scalar meson
$f_0$(975) (denoted hereafter as $\sigma^*$) and the vector meson $\phi$(1020) 
\citep{sch}.

The Lagrangian density (${\cal L}$) of the density-dependent RMF model is
given by \citep{hof1,hof2,bani02,typ10},
\begin{eqnarray}
{\cal L} &=& \sum_F \bar\Psi_{F}\left(i\gamma_\mu{\partial^\mu} - m_F
+ g_{\sigma F} \sigma - g_{\omega F} \gamma_\mu \omega^\mu 
- g_{\rho F} 
\gamma_\mu{\mbox{\boldmath $\tau$}}_F \cdot 
{\mbox{\boldmath $\rho$}}^\mu 
\right)\Psi_F\nonumber\\
&& + \frac{1}{2}\left( \partial_\mu \sigma\partial^\mu \sigma
- m_\sigma^2 \sigma^2\right) 
 -\frac{1}{4} \omega_{\mu\nu}\omega^{\mu\nu}\nonumber\\
&&+\frac{1}{2}m_\omega^2 \omega_\mu \omega^\mu
- \frac{1}{4}{\mbox {\boldmath $\rho$}}_{\mu\nu} \cdot
{\mbox {\boldmath $\rho$}}^{\mu\nu}
+ \frac{1}{2}m_\rho^2 {\mbox {\boldmath $\rho$}}_\mu \cdot
{\mbox {\boldmath $\rho$}}^\mu  + {\cal L}_{YY}~,
\label{lagm}
\end{eqnarray}
where $m_F$ is the bare mass of the baryon $F$ and 
${\mbox{\boldmath $\tau_{F}$}}$ is the isospin operator. 
Here $\Psi_F$ denotes the isospin multiplets for baryons. 
In principle, the sum may go over baryon multiplets 
F = N,$\Lambda$,$\Sigma$,$\Xi$. 

The appearance of hyperons depends on the hyperon-nucleon interaction strength 
in dense matter. The hyperon potential depths in normal nuclear matter are
determined from hypernuclei data \citep{sch,sch00,weis1,oertel12}. For example, 
the potential depth
of $\Lambda$ hyperons in nuclear matter at the saturation density is obtained
from $\Lambda$ hypernuclei data and is found to be attractive. 
Unlike $\Lambda$ hypernuclei 
data, $\Sigma$ hypernuclei data as well as $\Xi$ hypernuclei data are scarce.
This leads to large uncertainties in estimating the potential depths of 
$\Sigma$ and $\Xi$ hyperons in nuclear matter. It was noted that $\Sigma$
hypernuclei data indicated a repulsive $\Sigma$ potential depth in nuclear 
matter \citep{brat,sch00}. Such a repulsive $\Sigma N$ interaction might rule 
out the appearance of $\Sigma$ hyperons in dense matter 
or at least push their onset to very high densities. 
A few $\Xi$-hypernuclei data gave rise to a less attractive $\Xi$ potential 
depth in normal nuclear matter than the $\Lambda$ potential depth 
\citep{sch00,weis1,oertel12}.
Furthermore, $\Lambda$ hyperons, being the lightest hyperons among all hyperons,
would be populated first in the system
unless the potentials of the others would be very attractive. 
The appearance of heavier hyperons would
be delayed to higher densities.
Considering all these facts, we restrict ourselves to nucleons ($N$) and 
$\Lambda$ hyperons in this calculation. 
Despite this simplification, the EoS with $\Lambda$s included allows us to 
study the general features of strange degrees of freedom in core-collapse 
supernovae. Note that the same implication was used by \citet{peres}.

Arguments similar to those given above for the heavier hyperons apply for delta baryons, 
where it was typically found that these appear, if at all, only at the 
highest densities in neutron stars \citep{glendenning85}. In addition to the 
mass, the charge or isospin of hyperons and deltas is also important. 
It was only recently pointed out by \citet{pagliara14} 
that more modern density functionals that lead to lower symmetry energies
at high densities could give an earlier onset of deltas in neutron stars.
We leave the interesting aspect of including delta baryons for future study.

The Lagrangian density (${\cal L}_{YY}$)
responsible for hyperon-hyperon interaction is given by
\begin{eqnarray}
{\cal L}_{YY} &=& \sum_{F=\Lambda} \bar\psi_{F}\left(g_{\sigma^* F} \sigma^* 
- g_{\phi F} \gamma_\mu \phi^\mu
\right)\psi_F \nonumber\\
&& + \frac{1}{2}\left( \partial_\mu \sigma^*\partial^\mu \sigma^*
- m_{\sigma^*}^2 \sigma^{*2}\right)  \nonumber\\
&& -\frac{1}{4} \phi_{\mu\nu}\phi^{\mu\nu}
+\frac{1}{2}m_\phi^2 \phi_\mu \phi^\mu~.
\label{Lag}
\end{eqnarray}

The attractive $\Lambda-\Lambda$ interaction is mediated by the 
exchange of $\sigma^*$ meson. 
However, it is evident from double $\Lambda$ hypernuclei data that this 
attractive interaction is very weak \citep{taka,nakaza,gal}. Consequently, we 
omit the 
inclusion of $\sigma^*$ in Equation~({\ref{Lag}}). 

The field strength tensors for vector mesons are given by
\begin{eqnarray}
\omega^{\mu \nu}&=& \partial^ \mu \omega^ \nu-\partial^\nu \omega^ \mu 
\nonumber \\
\rho^{\mu \nu}&=& \partial^ \mu \rho^ \nu-\partial^\nu \rho^ \mu 
\nonumber \\
\phi^{\mu \nu}&=& \partial^ \mu \phi^ \nu-\partial^\nu \phi^ \mu. 
\nonumber \\
\end{eqnarray}

Though the structure of the density-dependent RMF Lagrangian density closely
follows that of the RMF model \citep{shen}, there are important differences 
between those models. In the RMF calculation with density-independent 
meson-baryon coupling 
constants, non-linear self interaction terms for scalar and vector fields 
are inserted to account for higher order density-dependent contributions. 
However, this is not necessary here as meson-baryon vertices $g_{\alpha F}$, where 
$\alpha$ denotes the $\sigma$, $\omega$ and $\rho$ fields, are dependent 
on Lorentz scalar functionals of baryon 
field operators and adjusted to the Dirac-Brueckner-Hartree-Fock (DBHF) 
calculations of nuclear matter \citep{typ99,hof1,hof2}. 

In mean field approximations adopted here, meson fields are replaced 
by their expectation values. Only the time-like components
of vector fields, and the third isospin component of $\rho$ 
fields have non-vanishing values in a uniform and static matter.
The mean meson fields are denoted by $\sigma$,  $\omega_0$,
$\rho_{03}$, and $\phi_0$. 

The grand-canonical thermodynamic potential per unit volume of the
 hadronic phase is given by 
\begin{eqnarray}
\frac{\Omega}{V} &=& \frac{1}{2}m_\sigma^2 \sigma^2
- \frac{1}{2} m_\omega^2 \omega_0^2 
- \frac{1}{2} m_\rho^2 \rho_{03}^2  
- \frac{1}{2} m_\phi^2 \phi_0^2 
- \Sigma^r \sum_{i=n,p,\Lambda} n_i
\nonumber \\
&& - 2T \sum_{i=n,p,\Lambda} \int \frac{d^3 k}{(2\pi)^3} 
[\mathrm{ln}(1 + e^{-\beta(E^* - \nu_i)}) +
\mathrm{ln}(1 + e^{-\beta(E^* + \nu_i)})] ~,  
\end{eqnarray}
where the temperature is defined as $\beta = 1/T$ and 
$E^* = \sqrt{(k^2 + m_i^{*2})}$.
In the present work, all Fermi-Dirac 
integrals are solved with the very accurate and efficient methods of 
\citet{aparicio98,gong01}, complemented by analytic approximations where 
these are even more reliable.

The chemical potential of $i$th baryon ($\mu_i$) is defined as 
\begin{equation}
\mu_{i} = \nu_i + \Sigma_i^v, 
\end{equation}
where $\Sigma_i^v$ is the vector self-energy and it is given by
\begin{equation}
\Sigma_i^v = g_{\omega i} \omega_0 + g_{\rho i} \tau_{3i} \rho_{03} + g_{\phi i}
\phi_0 + \Sigma^r~,
\label{vec}
\end{equation}
and the rearrangement term has the form
\begin{equation}
\Sigma^r=\sum_{i=n,p,\Lambda}[-\frac {\partial g_{\sigma i}} {\partial n_i} 
\sigma n_i^s 
+ \frac { \partial g_{\omega i}}{\partial n_i} \omega_0 n_i
+ \frac { \partial g_{\rho i}}{\partial n_i}\tau_{3i} 
\rho_{03} n_i
+ \frac { \partial g_{\phi i}}{\partial n_i} 
\phi_0 n_i]~.
\label{rea}
\end{equation}
Similarly, the expression of scalar self energy for $i$th baryon 
is given by 
\begin{equation}
\Sigma_i^s=g_{\sigma i}\sigma 
\label{sca}
\end{equation}
Now one can define the effective Dirac baryon mass as $m_i^*=m_i-\Sigma_i^s$. 

For our calculations we assume $\mu_{n} = \mu_{\Lambda}$, i.e., that there
is equilibrium with respect to strangeness changing reactions. This is 
justified because of the moderately long dynamic timescales in supernovae in 
the range of milliseconds, the high temperatures encountered inside the proto-neutron 
star, and because we expect that the $\Lambda$ hyperon abundance is only
significant at high densities, where weak equilibrium is established. 

Next we calculate the thermodynamic quantities of the baryonic matter such as
the pressure $P = - {\Omega}/V$ and the energy density 
\begin{eqnarray}
\epsilon &=& \frac{1}{2}m_\sigma^2 \sigma^2
+ \frac{1}{2} m_\omega^2 \omega_0^2 
+ \frac{1}{2} m_\rho^2 \rho_{03}^2  
+ \frac{1}{2} m_\phi^2 \phi_0^2 
\nonumber \\
&& + 2 \sum_{i=n,p,\Lambda} \int \frac{d^3 k}{(2\pi)^3} E^* 
\left({\frac{1}{e^{\beta(E^*-\nu_i)} 
+ 1}} + {\frac{1}{e^{\beta(E^*+\nu_i)} + 1}}\right)~.  
\end{eqnarray}
Similarly we can compute neutron, proton, and $\Lambda$ number densities which 
include contributions from both particle and antiparticles \citep{shen11}.  
The number density of the $i(=n,p,\Lambda$)-th baryon is 
$n_i = 2 \int \frac{d^3k}{(2\pi)^3} \left({\frac{1}{e^{\beta(E^*-\nu_i)} 
+ 1}} - {\frac{1}{e^{\beta(E^*+\nu_i)} + 1}}\right)$. 
The scalar density for the $i$th baryon ($n_i^s$) is 
\begin{eqnarray}
n_i^s &=& 
2 \int \frac{d^3 k}{(2\pi)^3} \frac{m_i^*}{E^*} 
\left({\frac{1}{e^{\beta(E^*-\nu_i)} 
+ 1}} + {\frac{1}{e^{\beta(E^*+\nu_i)} + 1}}\right) ~.
\end{eqnarray}
We can calculate the entropy density using 
${s} = \beta \left(\epsilon + P - \sum_{i=n,p,\Lambda} \mu_i n_i \right)$. 
The entropy per baryon is given by $S = {s}/{n_B}$, where $n_B$ is
the total baryon density
i.e., $n_B=\sum_i n_i$. 

The density dependence of nucleon-meson couplings was determined 
by \citet{typ99,typ10}. The functional forms of the density-dependent
couplings $g_{\sigma N}$ and $g{\omega N}$ are given by
\begin{eqnarray}
g_{\alpha N} = g_{\alpha N} (n_0) f_{\alpha} (x)~,\nonumber\\ 
f_{\alpha} (n_B/n_0) = a_{\alpha} \frac{1+b_{\alpha}(x+d_{\alpha})^2}
{1+c_{\alpha} (x +d_{\alpha})^2}~,
\label{coef}
\end{eqnarray} 
where $n_0$ is the saturation density, $\alpha = \sigma, \omega$ and 
$x = n_B/n_0$. For $\rho$ mesons, we have
\begin{eqnarray}
g_{\rho N} = g_{\rho N} (n_0) exp{[-a_{\rho} (x - 1)]}~. 
\label{rhoc}
\end{eqnarray} 

In this work we employ the DD2 parameter set \citep{typ10,fis2}, where the
coefficients in Equations (\ref{coef}) and (\ref{rhoc}), the saturation density,
the nucleon-meson couplings at the saturation density, and the 
mass of $\sigma$ mesons are determined by fitting the properties of finite
nuclei such as binding energies, spin-orbit splittings, charge and diffraction
radii, surface thickness, and neutron skin. In this fitting, experimental 
masses are used for the nucleons. 
In our EoS calculations, we also use the experimentally measured masses
of nucleons. 
The saturation properties of symmetric nuclear matter are 
obtained as $n_0 = 0.149065$ fm$^{-3}$, binding energy per nucleon 16.02 MeV, 
incompressibility 242.7 MeV, neutron effective Dirac mass 
$m_n^*/m_n$ = 0.5628, proton effective Dirac mass $m_p^*/m_p$ = 0.5622,
and the
symmetry energy 31.67 MeV. 
The value of the parameter corresponding to the density 
dependence of the symmetry energy at the saturation density is found to be 
55.03 MeV. 
For detailed definitions of these quantities, see e.g., \citet{composemanual}.
These nuclear matter properties are consistent with constraints
from theoretical 
calculations of neutron matter, experimental findings and astrophysical 
observations of neutron stars \citep{fis2,jim}.  
Note that the values that we obtain differ slightly from those previously 
reported in \citet{typ10}.
Meson-nucleon couplings at the 
saturation density and masses of baryons and mesons used in the calculation
are shown in Tables \ref{table1} and \ref{table2}, respectively.       

Nucleons do not couple with $\phi$ mesons i.e. $g_{\phi N} = 0$. 
The density-dependent 
meson-$\Lambda$ hyperon vertices are obtained from the density-dependent 
meson-nucleon couplings using $\Lambda$-hypernuclei data \citep{sch} and 
the SU(6) symmetry of the quark model.  
In the RMF model, vector meson - hyperon coupling constants were determined 
from the SU(6) symmetry relations of the 
quark model \citep{sch,dov}.  Similarly, we obtain
 vector meson - $\Lambda$ hyperon 
couplings in this model from the SU(6) symmetry 
relations \citep{sch}
\begin{eqnarray}
\frac{1}{2}g_{\omega \Lambda} =  \frac{1}{3} g_{\omega N},\nonumber\\
g_{\rho \Lambda} = 0, \nonumber\\
2 g_{\phi \Lambda} = - \frac{2 \sqrt{2}}{3} g_{\omega N}~. 
\end{eqnarray}

Next we obtain the scalar meson coupling to $\Lambda$ hyperons 
($g_{\sigma \Lambda}$) from the potential depth of $\Lambda$ hyperons in normal
nuclear matter. The $\Lambda$ hyperon potential in saturated 
nuclear matter is obtained from the experimental data of the single particle
spectra of $\Lambda$ hypernuclei. In the density-dependent RMF model, the 
potential depth of $\Lambda$ hyperon in saturated nuclear matter is given by 
\begin{equation}
U_{\Lambda}^N (n_0)=g_{\omega \Lambda} \omega_0 + \Sigma_N^r 
- g_{\sigma \Lambda} \sigma_0~ 
\label{pot}
\end{equation}
where $\Sigma_N^r$ is the contribution of only nucleons in the rearrangement 
term as given by Equation~(\ref{rea}).
In this calculation, the value of the $\Lambda$ potential in normal nuclear matter 
is taken as $U_{\Lambda} (n_0) = -30$ MeV \citep{mil,mar,sch92} and the 
ratio of $g_{\sigma \Lambda}$ and $g_{\sigma N}$ is 
$R_{\sigma \Lambda} = g_{\sigma \Lambda}/g_{\sigma N} = 0.62008$.  
Note that \citet{last} also extended the DD2 parameterization by including 
hyperons to describe the structures of hybrid stars. 
However, they used different assumptions, namely, an SU(3) rescaling with an 
overall factor $R=0.83$ and they considered the whole baryon octet.

\subsection{Extended Nuclear Statistical Equilibrium Model}  
\label{nse}
In the widely used nuclear EoS of Shen and
collaborators \citep{shen,shen11}, heavy nuclei were treated in the 
Thomas-Fermi approach. 
The other commonly used nuclear EoS of LS \citep{ls} 
utilizes a liquid-drop description of nuclei and a non-relativistic 
parameterization of the nucleon interactions. In both approaches
the gas of $\alpha$ particles was dealt with the Maxwell-Boltzmann 
statistics. Heavy nuclei are populated at low temperature and low density.
In the LS and Shen EoS, they used the single nucleus approximation
for heavy nuclei having an average representative atomic mass and charge in
inhomogeneous nuclear matter.

We adopt the extended NSE model of \citet{hs1} to describe the matter 
composed
of light and heavy nuclei along with unbound nucleons at low temperatures 
($\sim$ 10 MeV) and low densities below the saturation density. The region
where heavy nuclear clusters co-exist with nucleons is known as non-uniform
or inhomogeneous nuclear matter. 
In the HS model, nuclei are described as non-relativistic
particles using Maxwell-Boltzmann statistics and medium corrections
such as internal excitations or Coulomb screening. Excluded volume effects are
taken into account, which ensure the dissolution of heavy nuclei at high 
densities.
Interactions among unbound nucleons are described by Equation (\ref{lagm}),
employing the same parameter set DD2, but not including hyperons. This is 
justified because the fraction of $\Lambda$ hyperons is 
negligibly small in low-temperature and -density domains. 

Approximately 8000 nuclear species 
are considered in the extended NSE model of HS. 
Experimental masses of nuclei ($A \geq 2$) used in the model are taken 
from the atomic mass table of \citet{audi}. For exotic nuclei without
measured masses, theoretical nuclear structure calculations within the 
framework of finite-range droplet model (FRDM) \citep{moel} are exploited.
Note that nuclei beyond the neutron drip line are not considered. By
using nuclear mass tables, 
nuclear shell effects are automatically included into the calculation. 
This is necessary to obtain the correct low-density limit, e.g. relevant
for consistency with recent electron-capture rates (see \citet{juodagalvis10})
and with the simulation of the progenitor star, or if one wants to connect to a 
non-NSE EoS. However, we also point out that the modification of the 
nuclear shell structure at high densities is not well described by the HS approach.
The HS EoS goes beyond the single nucleus approximation (SNA). 
Regarding a distribution of only heavy nuclei, it is well known that the SNA
has only a small effect on thermodynamic quantities \citep{burrows84}. However, here we 
also include various light nuclei, which together with unbound nucleons dominate 
the composition of shock-heated matter \citep{sumi08} and have a non-negligible impact 
on thermodynamic quantities \citep{hs1}. Note that the results for light nuclei of the HS model
are in good agreement with that of the quantum many-body calculation
\citep{hemp} and also qualitatively with experimental data from 
heavy-ion collisions \citep{qin12}. 

The thermodynamic quantities such as pressure, energy density, etc., are 
obtained from the total canonical partition function given by
\begin{equation}
Z(T,V,\{N_i\})=Z_{\rm nuc}~\prod_{A,Z}Z_{A,Z}~Z_{\rm Coul} \; ,
\end{equation}
with $V$ denoting the volume of the system. 
One can write down the Helmholtz free energy using the partition function as,
\begin{eqnarray}
F(T,V,\{N_i\})&=&-T \mathrm{ln} Z \\
&=& F_{\rm nuc}+\sum_{A,Z} F_{A,Z} +F_{\rm Coul} \; ,
\end{eqnarray}
where $F_{\rm nuc}$, $F_{\rm Coul}$, $F_{A,Z}$ are the free energies of 
nucleons, the
Coulomb free energy, and the free energy of the nucleus
represented by the Maxwell-Boltzmann distribution \citep{hs1}. 

After implementing the excluded volume effects in a thermodynamically consistent
manner, the number density of the nuclei is given by \citep{hs1}
\begin{eqnarray}
&&n_{A,Z}=\kappa~g_{A,Z}(T)\left(\frac{M_{A,Z} T}{2\pi}\right)^{3/2}\exp\left(\frac{(A-Z)\mu_{n}^0+Z\mu_{p}^0-M_{A,Z}-E^{\rm Coul}_{A,Z}-P^0_{\rm nuc}V_{A,Z}}T\right) \; , \label{eq_naz}
\end{eqnarray}
where $\kappa$ is the volume fraction available for nuclei and defined in 
terms of local number 
densities and takes values between 0 and 1. It may be worth noting that 
Equation ({\ref{eq_naz}}) can be used to derive a 
modified Saha equation due to excluded volume corrections. 

Next one can define the free energy density \citep{hs1}
\begin{eqnarray}
f&=&\sum_{A,Z} f_{A,Z}^0(T,n_{A,Z})+f_{\rm Coul}(n_e,n_{A,Z})+\xi f_{\rm nuc}^0(T,n'_n,n'_p)-T\sum_{A,Z} n_{A,Z} \mathrm{ln}(\kappa)\; ,
\label{fe}
\end{eqnarray}
where the first term is the contribution of the non-interacting gas of 
nuclei. Here $f_{\rm Coul}$ is the Coulomb free energy.
The free energy density of the interacting nucleons $f_{\rm nuc}^0$ 
is multiplied by the available volume fraction of nucleons $\xi$. 
The local number densities of neutrons and protons 
are denoted by $n'_n$ and $n'_p$, respectively. 
The last term, which corresponds to a hard-core repulsion of nuclei, 
goes to infinity when $\kappa$ approaches zero near saturation density and the
uniform matter is formed.

The energy density is given by the following expression \citep{hs1}:
\begin{eqnarray}
\epsilon&=&\xi \epsilon_{\rm nuc}^0(T,n'_n,n'_p)+\sum_{A,Z} \epsilon_{A,Z}^0(T,n_{A,Z})+f_{\rm Coul}(n_e,n_{A,Z})\; ,
\label{enint}
\end{eqnarray}
\begin{eqnarray}
\epsilon_{A,Z}^0(T,n_{A,Z})&=& n_{A,Z}\left(M_{A,Z}+\frac32T+\frac{\partial g}{\partial T}\frac {T^2} g\right)\; .
\end{eqnarray}
Similarly the total pressure becomes
\begin{eqnarray}
P&=&P_{\rm nuc}^0(T,n'_n,n'_p)+\frac1{\kappa}\sum_{A,Z} P_{A,Z}^0(T,n_{A,Z})+ P_{\rm Coul}(n_e,n_{A,Z})~, \label{eq_p}
\end{eqnarray}
\begin{eqnarray}
P_{A,Z}^0(T,n_{A,Z})&=&Tn_{A,Z}~.
\end{eqnarray}

Note that all quantities above relating to nucleon contributions are calculated
with the RMF model (DD2), as described in Section~\ref{rmf}
and taking into account general Fermi-Dirac statistics. In the 
original work of \citet{hs1}, TMA interactions were used instead. Some further
changes were made to improve the description of non-uniform matter. Here we briefly list only
the relevant ones: for simplicity, nuclei are only considered up to a 
temperature of 50 MeV, instead of 20 MeV used previously. In the internal partition
function of nuclei, $g_{A,Z}(T)$ in Equation~(\ref{eq_naz}), which is taken from
\citet{fai82}, only excited states
up to the binding energy of the corresponding nucleus are included. 
Our basic idea is that we want to keep the nucleus bound.
If no cutoff in the integral for the excited states was used, 
arbitrarily large excitation energies would contribute
to the energy density. The energy and entropy stored in nuclei would 
increase with increasing temperature to unphysically large values. 
We found in different applications of the EOS that the usage of the cutoff 
leads to a more well-balanced behavior. Nevertheless, it is clear that our 
description of excited states remains on a rather heuristical level.

\subsection{Matching Procedure}
\label{match}
In principle, the hyperonic EoS as presented in Section~\ref{rmf} could be 
used directly for the description of the unbound baryon contribution 
(denoted by the subscript ``nuc'')
in the statistical model that was summarized in Section~\ref{nse}. However, 
because a complete nucleonic supernova EoS table for the parameter set DD2 
and as described in Section~\ref{nse} is already publicly 
available, called HS(DD2) (see \citet{fis2}), here we follow a different 
strategy. We expect that nuclei will not be present with high abundances 
at conditions where hyperons can be formed, i.e., at high densities or high
temperatures. Therefore, we do not repeat the calculation of the EoS 
with non-uniform matter distributions including hyperons, 
but only replace certain parts of the existing table
with the new uniform hyperonic EoS 
using physical criteria specified in the following. In consequence, the new 
tables never 
contain a mixture of hyperons and nuclei. 

For the merging of the two tables, we follow a
standard thermodynamic criterion, namely that the free energy per baryon at 
fixed $T$, $n_B$, and $Y_p$ has to be minimized. However, this physical 
criteria alone could lead to odd transition behaviors, because transitions
from one EoS to the other could also be induced by numerical errors.
Here, such unphysical transitions are avoided by introducing a minimal
hyperon mass fraction of 10$^{-5}$, i.e., the hyperon EoS replaces the 
nucleonic EoS only if it has a lower free energy per baryon and if 
$X_\Lambda >10^{-5}$. Using these two criteria for the merging of the two EoS,
we obtain a smooth and continuous transition boundary.

\section{Results and Discussion}       
We compute the hyperon EoS tables using the DD2 parameter set of 
Table \ref{table1}. We denote the hyperon EoS table without $\phi$ mesons as 
BHB$\Lambda$ corresponding to the composition np$\Lambda$ 
and the hyperon EoS table with $\phi$ mesons as BHB$\Lambda \phi$
for the np$\Lambda\phi$ case. 
In both cases, the tables are constructed for 
temperatures $T=0.1$ to $10^{2.2}\simeq 158.49$~MeV and proton fractions $Y_p$ = 0.01 to 0.6, 
whereas baryon densities range from $n_B = 10^{-12}$ to 1 fm$^{-3}$ for 
the BHB$\Lambda$ and $n_B = 10^{-12}$ to $10^{1.08} \simeq 1.2$~fm$^{-3}$ for 
the BHB$\Lambda\phi$ tables. 
We have different density ranges for the two tables (which are also 
different compared to the original nucleonic HS(DD2) table), because we could 
not obtain physical solutions at higher values.
We adopt a linear grid spacing of
0.01 for $Y_p$ and logarithmic grid spacing of 0.04 for $T$ and $n_B$. 
An overview of the two EoS tables is given in Table \ref{tab:overview}. 
Before we go into a detailed description of thermodynamic quantities in 
hyperon EoS tables, we discuss the $\beta$-equilibrated matter relevant for 
cold neutron stars.

\begin{figure}
\epsscale{0.5}
\plotone{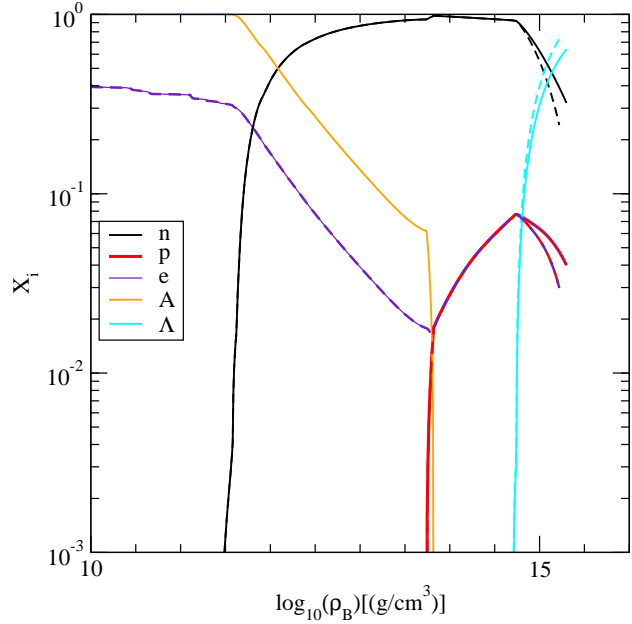}
\caption{Mass fractions of various species are plotted as a function of
baryon mass density with (solid lines) and without (dashed lines) $\phi$ mesons
in $\beta$-equilibrated hyperon matter. The curve labeled with ``A'' 
shows the mass fraction of heavy nuclei.} 
\label{fracb}
\end{figure}

\begin{figure}
\epsscale{0.53}
\plotone{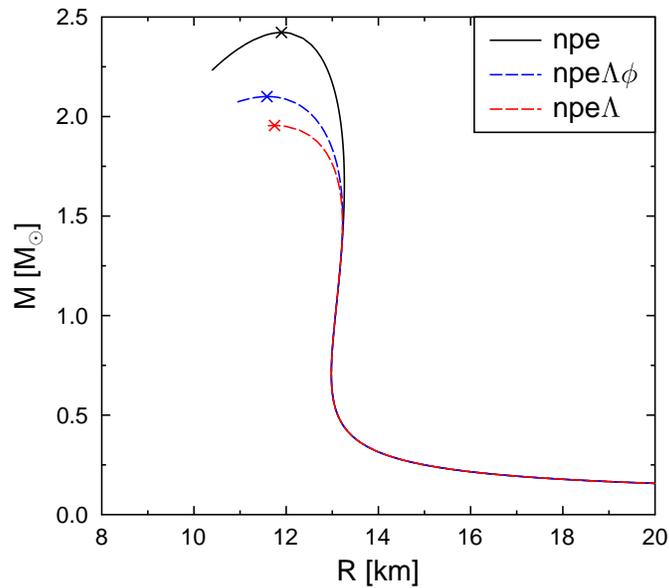}
\caption{Masses of the neutron star sequence are plotted as a function of
radius with (online-version: blue) and without (online-version: red) $\phi$ 
mesons in hyperon and nucleon matter (solid line), 
corresponding to the BHB$\Lambda \phi$, BHB$\Lambda $, and HS(DD2) EoS,
respectively. Crosses mark the maximum
mass configurations.} 
\label{mr}
\end{figure}

\begin{figure}
\epsscale{0.45}
\plotone{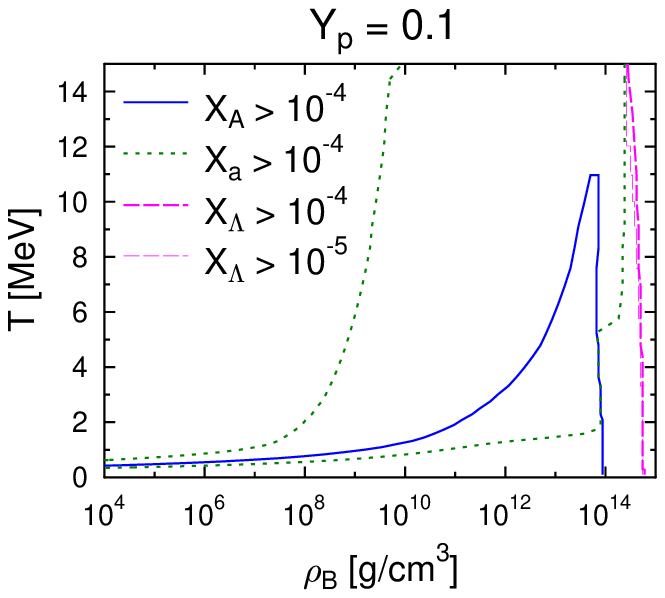}

\vspace{0.2cm}
\plotone{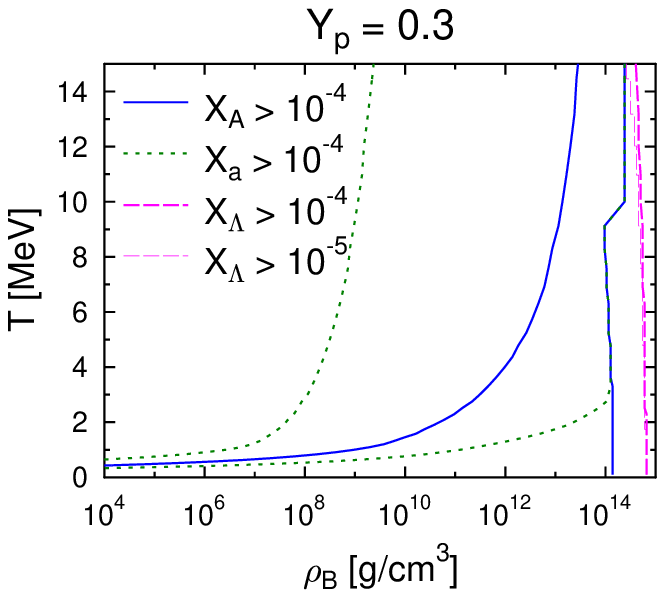}

\vspace{0.2cm}
\plotone{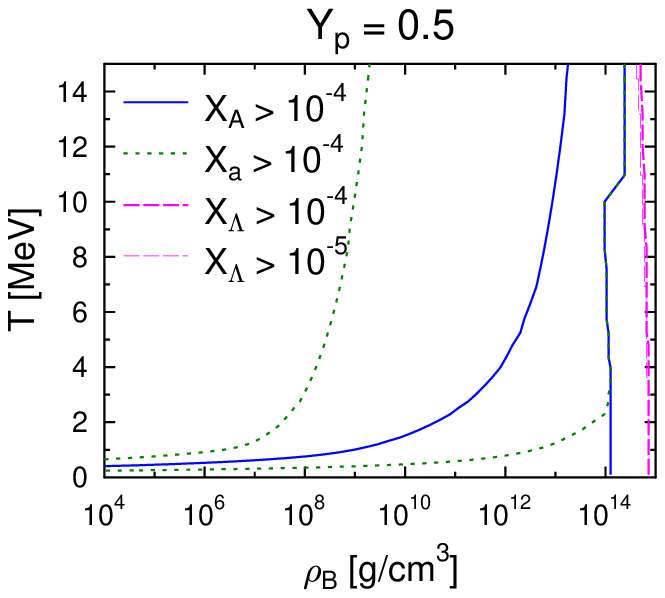}
\caption{Phase diagrams at $Y_p=0.1$, 0.3, and 0.5 (bottom to top) in the 
$T$-$\rho_B$ plane for the BHB$\Lambda \phi$ EoS. The lines delimit regions 
where the mass fractions of 
light nuclei ($X_a$), heavy nuclei ($X_A$) and $\Lambda$ hyperons ($X_\Lambda$) 
exceed $10^{-4}$. The thin dashed magenta line also
shows where the mass fraction of $\Lambda$s exceeds $10^{-5}$. 
$\Lambda$s occur abundantly only at high densities.}
\label{pd_yp}
\end{figure}

\begin{figure}
\epsscale{0.45}
\plotone{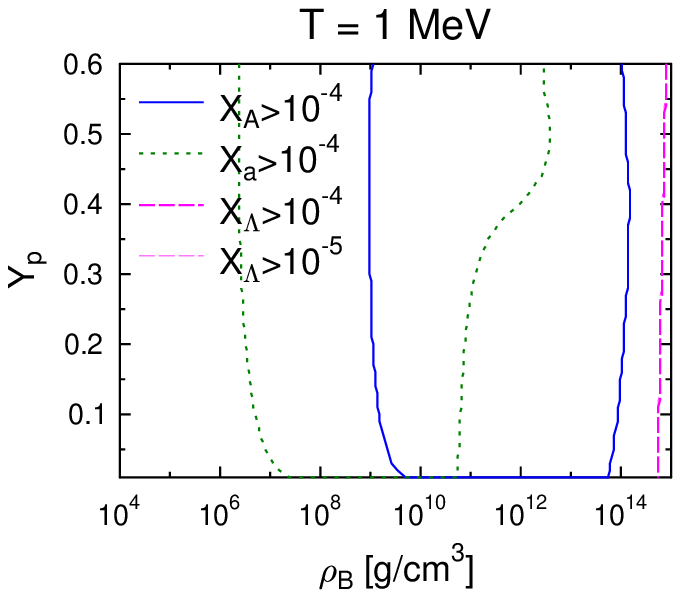}

\vspace{0.2cm}
\plotone{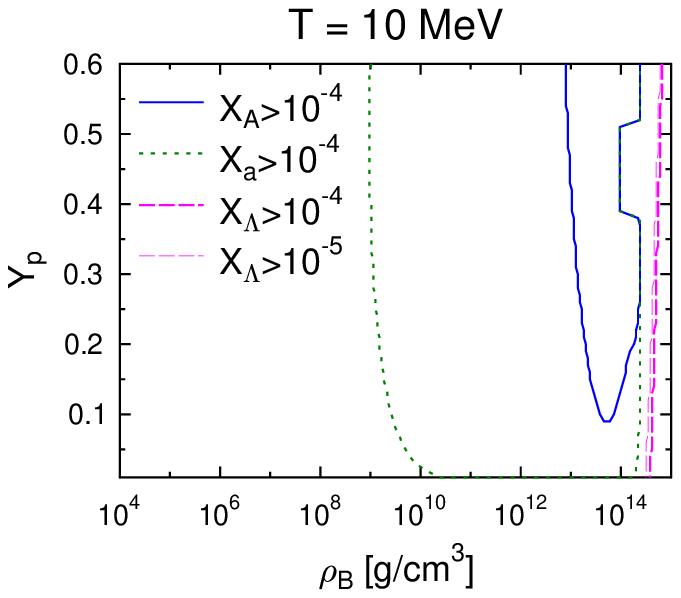}

\vspace{0.2cm}
\plotone{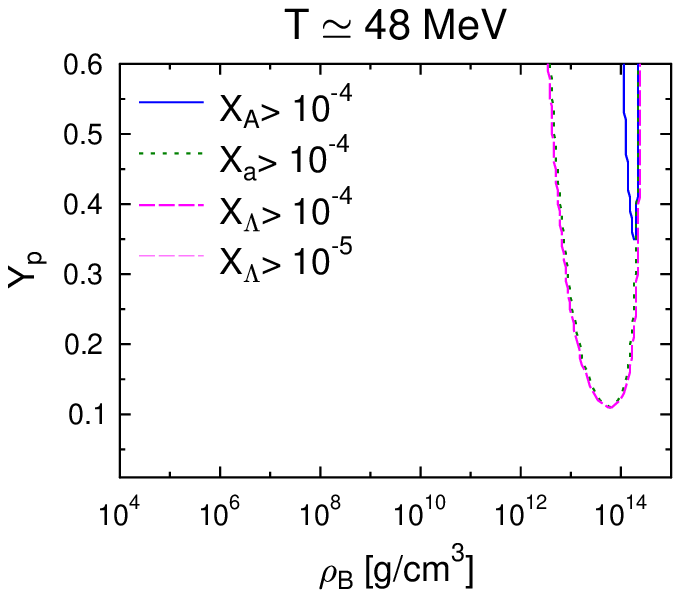}
\caption{Phase diagram at $T=1$, 10, and $\simeq48$~MeV (bottom to top) in the 
$Y_p$-$\rho_B$ plane for the BHB$\Lambda \phi$ EoS. 
The lines delimit regions where the mass fractions of 
light nuclei ($X_a$), heavy nuclei ($X_A$) and $\Lambda$s ($X_\Lambda$) 
exceed $10^{-4}$. The thin dashed magenta line also
shows where the mass fraction of $\Lambda$s exceeds $10^{-5}$.
For $T=1$ and 10~MeV, $\Lambda$s occur abundantly only at high densities. For
$T\simeq48$~MeV, the isocontours of light nuclei and $\Lambda$s almost 
coincide.} 
\label{pd_t}
\end{figure}

We generate the EoS of neutron stars by imposing charge neutrality with the 
inclusion of electrons and the $\beta$-equilibrium condition without neutrinos into
hyperon EoS tables at a very low temperature $T=0.1$ MeV. 
Fractions of 
different particle species in $\beta$-equilibrated hyperon matter with and
without $\phi$ mesons are plotted as a function of baryon mass density in 
Figure {\ref{fracb}. 
Here and in the following, we define a ``baryon mass density'' $\rho_B$ by
$\rho_B=n_B\cdot m_u$, with the atomic mass unit $m_u$.
The solid lines represent the npe$\Lambda\phi$ case whereas the dashed lines 
represent the npe$\Lambda$ case. The beginning of the inner
crust is clearly visible by the sudden appearance of free neutrons at a 
density of $\sim 3 \times 10^{11}$ g/cm$^3$.
In both cases, heavy nuclei dissolve into their 
fundamental constituents, i.e., nucleons, below the saturation density and a 
uniform nuclear matter is formed just after that,
marking the transition to the neutron star core.
There, proton fractions increase as baryon density increases. 
The positive charges of protons are 
balanced by negative charges of electrons. When the baryon density reaches
2.1~$n_0$,
$\Lambda$ hyperons begin to populate the system in both cases. As
the $\Lambda$ fraction rises, neutron and proton fractions drop. Furthermore,
it is noted that the $\Lambda$ fraction for the np$\Lambda$ case is higher than 
that of the 
np$\Lambda\phi$ case. This may be attributed to the strong repulsive interaction
due to $\phi$ mesons at higher densities. This might have a significant impact on 
the EoS and mass-radius relationship of neutron stars with and without $\phi$ 
mesons.    

The mass-radius
relationship of the sequence of neutron stars is shown in Figure {\ref{mr}}. 
The solid line represents the nucleons-only neutron star. On the other hand, bold 
(online-version: blue) and light dashed (online-version: red)
lines represent neutron stars including hyperons with and
without $\phi$ mesons, respectively. Note that the crust EoS is 
contained self-consistently, i.e., no external models have to be used.
It is evident from Figure~{\ref{mr}} that
$\Lambda$ hyperons make the EoS softer, resulting in a smaller maximum mass 
neutron star
compared with that of the nucleons-only case. Further, we find that the 
hyperon-hyperon interaction mediated by $\phi$ mesons makes the hyperon EoS
stiffer than the case without $\phi$ mesons. Consequently, 
the npe$\Lambda\phi$ case has a higher maximum mass than that
of the npe$\Lambda$ case because of the repulsive contribution of $\phi$ 
mesons in the hyperon EoS in the former case. The maximum masses corresponding to 
the nucleons-only, npe$\Lambda$, and npe$\Lambda\phi$ neutron star sequences are
2.42, 1.95 and 2.10 
M$_{\odot}$, respectively. 
We remark that the different extension of the neutron star DD2 EoS with 
hyperons 
done previously by \citet{last} gave a maximum mass of only 1.94~M$_\odot$.
It is important to note that the maximum mass of 
npe$\Lambda\phi$ case is well above the benchmark-measured neutron star mass of 
2.01$\pm$0.04 M$_{\odot}$ \citep{anto}. This is 
the first supernova EoS with 
hyperons that is compatible with a 2 M$_{\odot}$ neutron star. 

We calculate the strangeness fractions $f_s$ in maximum mass neutron
stars, defined as the ratios of the total numbers of strangeness and the total 
baryon numbers, and find that in the npe$\Lambda$ case it is 
0.071 and in the npe$\Lambda\phi$ case it is 0.059.
Using these values we can roughly confirm the empirical relation of 
\citet{weis2}, 
\begin{equation}
\frac{M_{\rm max}^{\rm emp}}{M_{\odot}} = \frac{M_{\rm max}(f_s=0)}{M_{\odot}} 
- c (\frac{f_s}{0.1})~,
\end{equation}
where $c=0.6$, and that the maximum mass reduces with the strangeness fraction. 

Figure {\ref{pd_yp}} gives a general overview of the composition. The 
lines delimit regions where the mass fractions of light and heavy nuclei, and 
of $\Lambda$ hyperons that exceed $10^{-4}$. 
Light and heavy nuclei are distinguished here via their charge number
($Z\leq 5$ and $Z\geq 6$, respectively). 
For $\Lambda$s in Figure {\ref{pd_yp}}, the 
minimal mass fraction of $10^{-5}$ is also shown, which marks the transition to 
np$\Lambda$-matter and thus shows the results of the matching procedure.
The structure of the regions where light and heavy nuclei 
are abundant is similar to what was reported in \cite{hs1}. Regarding 
$\Lambda$ hyperons, we observe that for the low temperatures selected in 
Figure {\ref{pd_yp}}, there is no overlap with the regions where nuclei appear.
For the conditions shown in Figure {\ref{pd_yp}},
$\Lambda$s instead only appear for densities above 
$ \sim n_0$ whereas their onset
is slightly decreasing with increasing temperature. We also observe that they 
are slightly more abundant for low $Y_p$. 

Figure {\ref{pd_t}} gives complementary information about the composition by
showing ``phase diagrams'' in the $Y_p$-$\rho_B$ space. For $T=10$~MeV, there 
is an unexpected kink for $X_A$ for $Y_p$ between 0.4 and 0.5. This is probably
related to the limitation of the composition in the HS(DD2) model regarding
the maximum asymmetry and mass number that nuclei can have. In the bottom 
panel of Figure {\ref{pd_t}}, a temperature of $\simeq 48$~MeV is selected.
Note again that for $T\geq50$~MeV, the HS model does not take into account the
formation of nuclei. The temperature of $\simeq 48$~MeV corresponds 
to the highest temperature in our final EoS tables, where nuclei can, 
in principle, still appear. 
It is important to note that the small fraction of ``heavy'' nuclei $X_A$ 
that can be seen in Figure  {\ref{pd_t}} at this high temperature, 
actually corresponds to intermediate mass nuclei 
such as carbon. Here the abundance of nuclei is decreasing exponentially with 
their mass number.
At the temperature of $\simeq 48$~MeV we find that the lines for $\Lambda$s 
almost coincide with those of light nuclei. For moderate asymmetries 
(e.g., $Y_p=0.3$),
an isothermal compression would lead to a transition from np$\Lambda$-matter
to a mixture of nucleons and light nuclei, and then above 
$\sim 10^{14}$~g/cm$^3$ back
to np$\Lambda$-matter. This ``peninsula'' of light nuclei must be seen as a 
result of the minimization of the free energy. For
very low densities and such high temperatures, there would be almost no nuclei 
present. Instead, there is a thermal contribution of $\Lambda$s that makes them
the favorite phase. At intermediate densities, light nuclei
play a more important role than $\Lambda$s. At high densities, where the
formation of $\Lambda$s is driven by density and by high 
chemical 
potentials, they again form the most stable phase. As mentioned earlier, a 
more detailed calculation should, in principle, consider all possible degrees of 
freedom at all conditions. 

Figure \ref{fracp}} exhibits the composition of supernova matter at different 
regimes of temperatures, proton fractions and densities. 
Fractions of neutrons ($X_n$), protons ($X_p$), light nuclei ($X_a$), 
heavy nuclei ($X_A$), and $\Lambda$s ($X_{\Lambda}$) are shown as
a function of baryon mass density for $T = 1$, 10 and 100 MeV and $Y_p =$ 0.1, 
0.3 and 0.5, for the np$\Lambda \phi$ case. 
For $T=1$~MeV and $Y_p =$ 0.1, almost only free neutrons and protons exist
up to a mass density of $\sim 10^{7}$ g/cm$^{3}$. Beyond this density point, the free
proton fraction drops sharply because protons are now bound inside light 
nuclei coexisting with free neutrons, which at this low temperature are mostly 
alpha-particles. Similarly, the free neutron fraction is
reduced. The shape of the curve for light nuclei tends to be symmetric and the 
width of it increases with higher values of proton fraction. 
Heavy nuclei ($Z\geq 6$)
start populating the system around $10^{9}$ g/cm$^{3}$ replacing light nuclei.
This trend is noted also for other values of proton fractions.
The fraction of heavy nuclei grows and reaches its maximum value at higher
mass densities with an increasing proton fraction as is evident from the $T=1$~MeV 
panel of Figure~{\ref{fracp}}. Consequently, fractions of free neutrons and light 
nuclei fall rapidly. Heavy nuclei dissolve into their fundamental constituents
at $\sim 10^{14}$ g/cm$^{3}$ and form a uniform matter of neutrons and protons.
It is observed that $\Lambda$ hyperons appear with significant abundance
at a density above 2$n_0$  
at the cost of neutrons when the zero-temperature threshold condition 
$\mu_\Lambda = \mu_n \geq  m_{\Lambda}$ is satisfied. 
The higher the proton fraction, the smaller the 
population of $\Lambda$s. Note again that we include $\Lambda$ hyperons in the
hyperon EoS tables only when its fraction is above $10^{-5}$. 

\begin{figure}
\epsscale{1.0}
\plotone{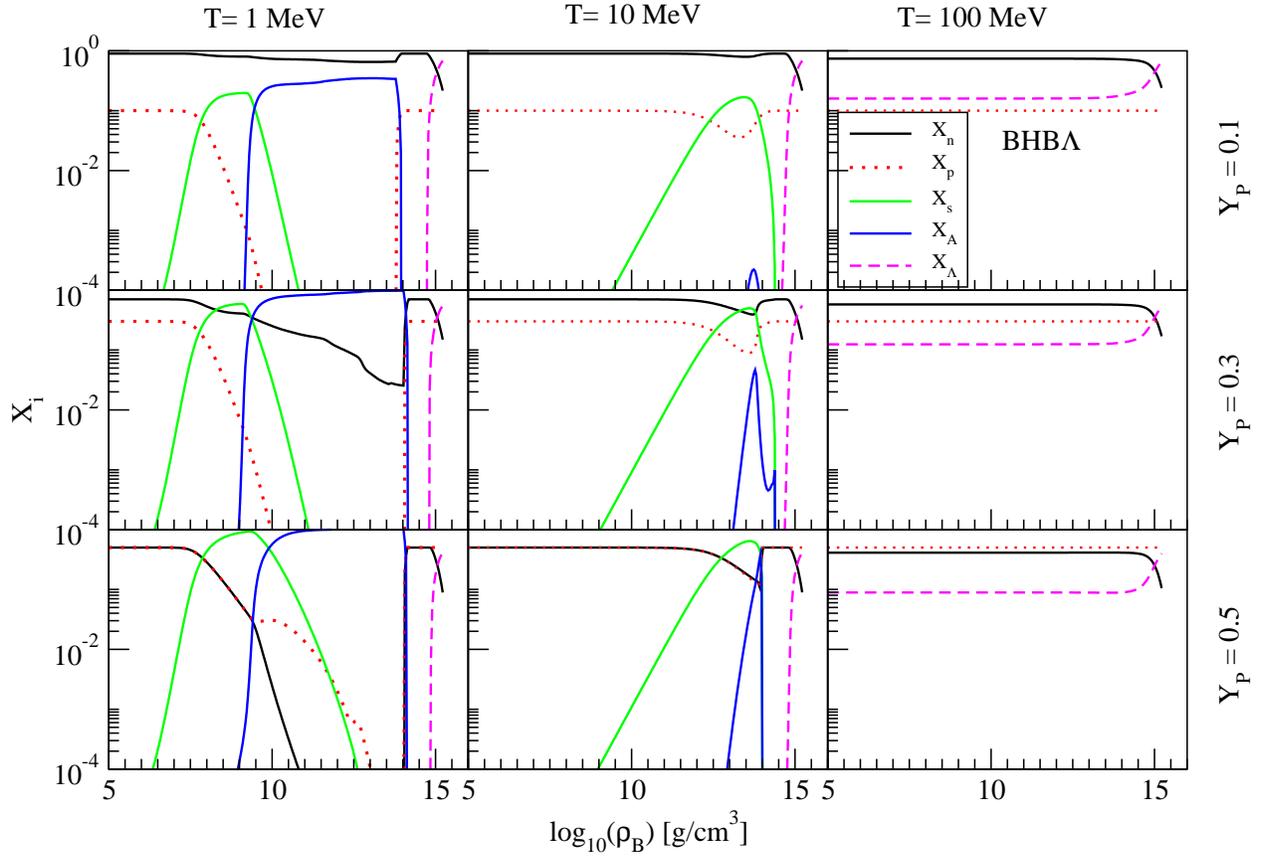}
\caption{ Mass fractions of neutrons ($X_n$), protons ($X_p$), light nuclei 
($X_a$), heavy nuclei ($X_A$), and $\Lambda$ hyperons ($X_{\Lambda}$) are 
plotted as a function of baryon mass density for BHB$\Lambda$
corresponding to the np$\Lambda$ case.
} 
\label{frac}
\end{figure}

\begin{figure}
\epsscale{1.0}
\plotone{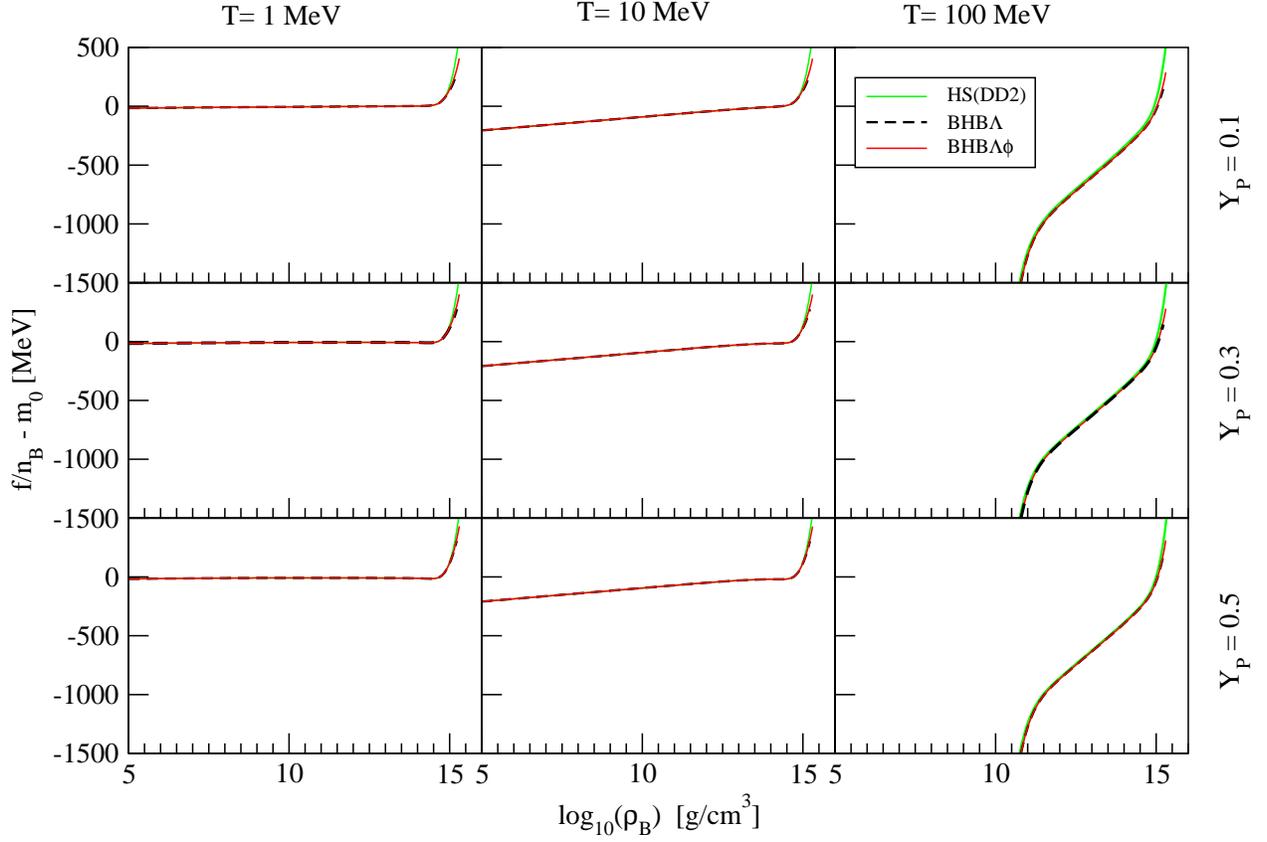}
\caption{Free energy  per baryon with respect to $m_0 = 938$ MeV is plotted
as a function of baryon mass density for temperatures $T=$1, 10, 100 MeV and
proton fractions $Y_p =$ 0.1, 0.3 and 0.5. 
Results from the nucleonic EoS table HS(DD2) (online-version: green) and the two 
hyperon EoS tables BHB$\Lambda$ (dashed line) and BHB$\Lambda \phi$ 
(online-version: red) are shown here.
}
\label{fen}
\end{figure}

\begin{figure}
\epsscale{1.0}
\plotone{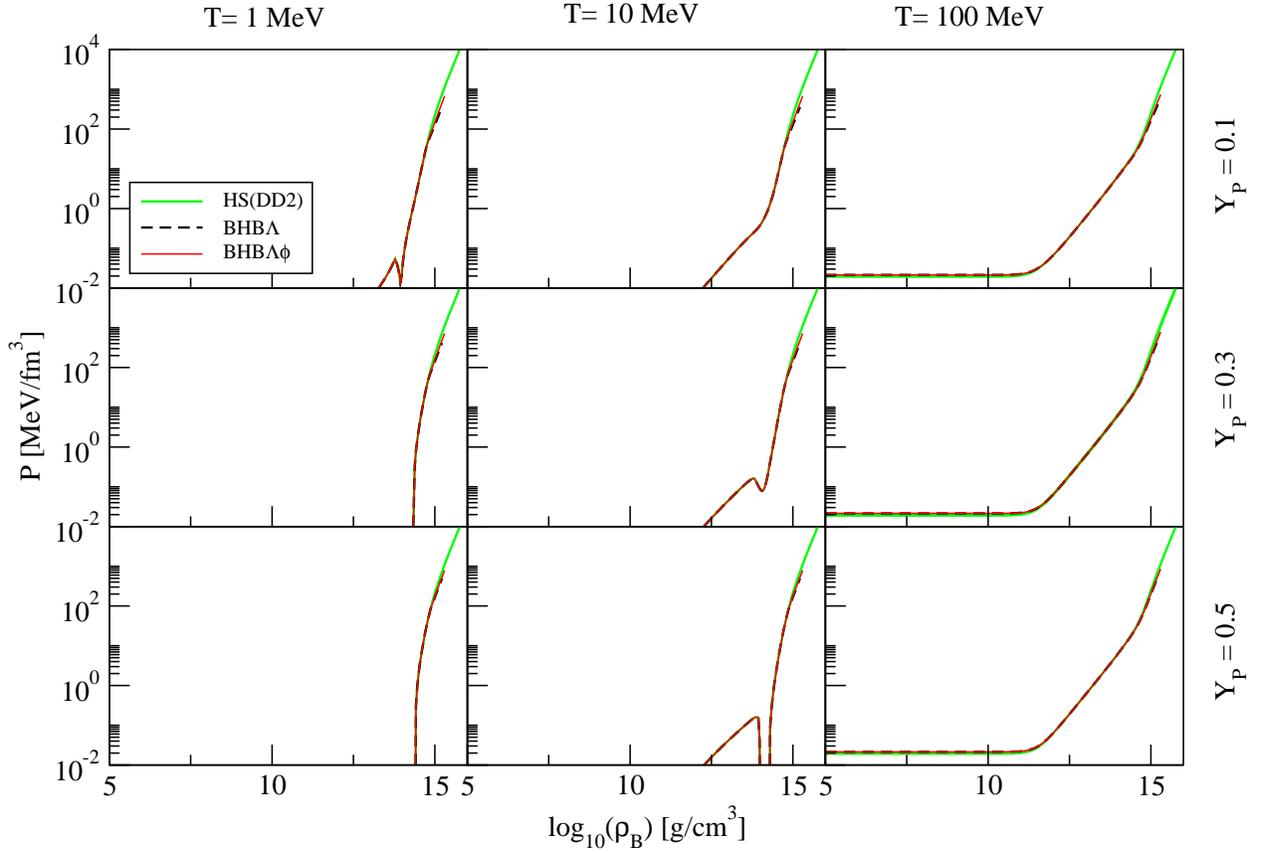}
\caption{Pressure is exhibited
as a function of baryon mass density for temperatures $T=$1, 10, 100 MeV and
proton fractions $Y_p =$ 0.1, 0.3 and 0.5.
Results from the nucleonic EoS table HS(DD2) (online-version: green) and the two 
hyperon EoS tables BHB$\Lambda$ (dashed line) and BHB$\Lambda \phi$ 
(online-version: red) are shown here.
}
\label{pres}
\end{figure}

\begin{figure}
\epsscale{1.0}
\plotone{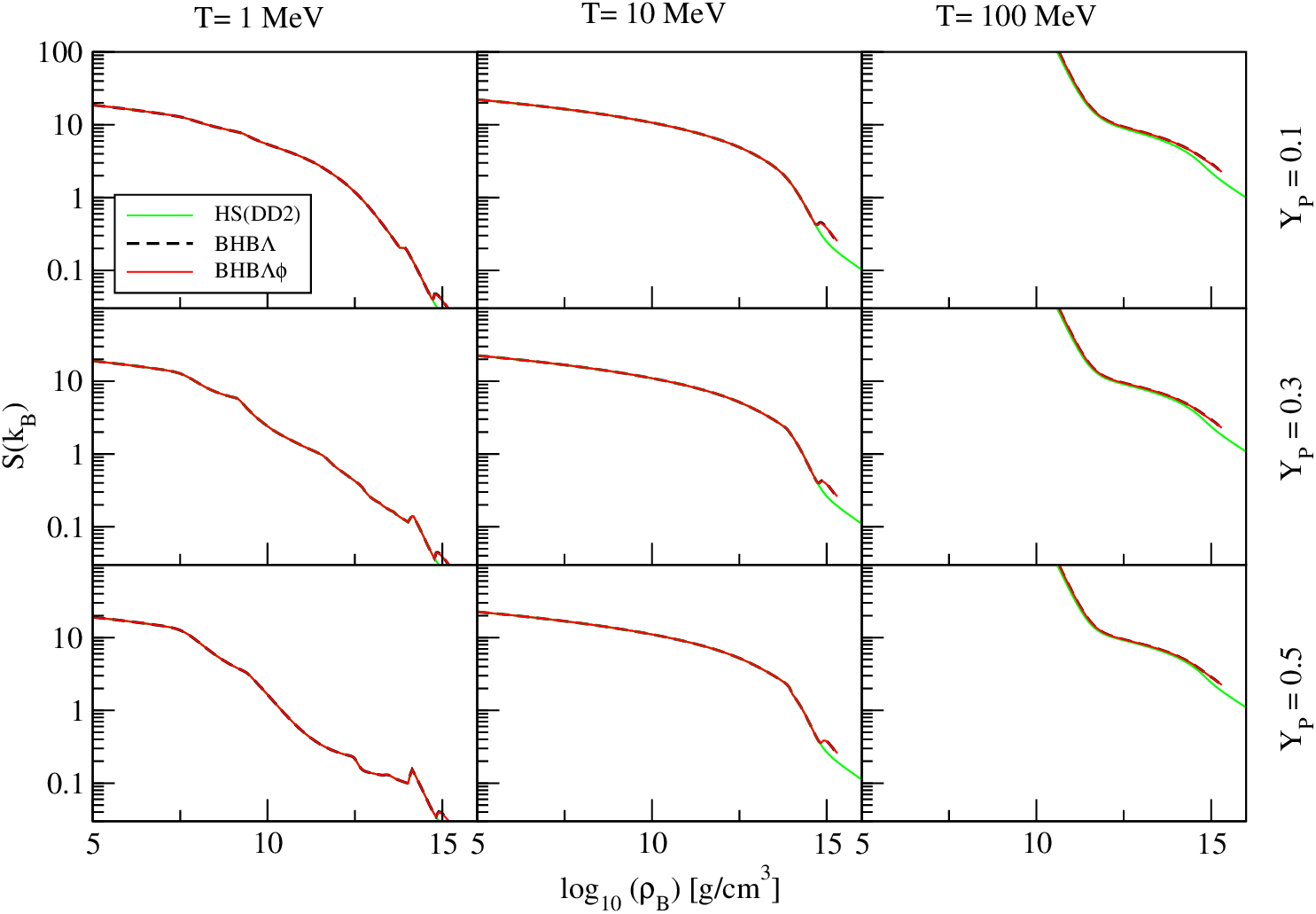}
\rotate
\caption{Entropy per baryon is shown 
as a function of baryon mass density for temperatures $T=$1, 10, 100 MeV and
proton fractions $Y_p =$ 0.1, 0.3 and 0.5.
Results from the nucleonic EoS table HS(DD2) (online-version: green) and the two 
hyperon EoS tables BHB$\Lambda$ (dashed line) and BHB$\Lambda \phi$ 
(online-version: red) are shown here.
} 
\label{entp}
\end{figure}

Now we focus on the case of $T=10$~MeV of Figure~{\ref{fracp} (middle panel). Here
light nuclei are formed replacing free nucleons at higher mass densities. 
Though significant populations of light nuclei are noted for different $Y_p$,
the distribution of heavy nuclei is appreciable and very sharp only for 
$Y_p =$ 0.5. It was found that just like the $T=1$~MeV case, nuclei melt down to form a uniform nuclear
matter before the saturation density is reached and $\Lambda$s appear at higher
densities. For $T=1$ and 10~MeV, we do not find any thermal $\Lambda$s as 
expected.

Next we discuss the case of $T=100$~MeV in Figure~{\ref{fracp}} (right panel). 
Note that in the HS(DD2)-EoS nuclei are only considered up to a temperature of
50~MeV, because their contribution is small for such 
high temperatures. Thus, only uniform matter of
neutrons, protons and $\Lambda$s is found to exist in this case. A significant
fraction of $\Lambda$ hyperons is thermally produced at low densities with the 
constraint $\mu_n = \mu_{\Lambda}$ and it grows with density at the expense of
neutrons. We find qualitatively similar behavior for the hyperon case without
$\phi$ mesons as shown in Figure~{\ref{frac}}. The only difference between 
the np$\Lambda$ and np$\Lambda \phi$
cases is found in 
$\Lambda$ fraction at high densities as is also evident from 
Tables {\ref{table3}} - {\ref{table6}} which show excerpts from our full
EoS tables. 

These tables also show that very low values of the effective Dirac masses are
found at very high densities. Small or even negative values of the nucleon 
effective Dirac mass are well known to occur in relativistic mean-field models 
(see \citet{zimanyi90,sch}). In our case with $T=0.1$~MeV, 
vanishingly small or negative values of the nucleon effective Dirac mass occur beyond the
central density, corresponding to the maximum neutron star mass.   

Free energy per baryon is shown as a function of baryon mass density in 
Figure {\ref{fen}}. 
Here and in all following plots, we only show the baryonic contribution.
Free energy per baryon is measured with respect to 
the arbitrary value of
$m_0 = 938$ MeV. This figure also shows various regimes of 
temperatures, $T=$ 1, 10 and 100 MeV, and proton fractions, $Y_p=$ 0.1, 0.3 and 
0.5. Furthermore, the results of hyperon matter with (solid line) and without 
(dashed line) $\phi$ mesons are shown in Figure~{\ref{fen}}. At lower densities,
there is practically no difference between the results of 
nuclear and hyperon 
matter for different situations considered. This may be attributed to 
no $\Lambda$s for $T=$ 1 and 10 MeV or just a low abundance of thermal 
$\Lambda$s in the case of $T=$ 100 MeV as shown in the previous figures. 
On the other hand, the free energy is 
reduced when
$\Lambda$s are populated significantly at higher densities and higher 
temperatures compared with the nuclear EoS. It is noted that when 
hyperon-hyperon interaction is mediated by $\phi$ mesons 
in the np$\Lambda \phi$ case, the free energy is higher 
than that of the np$\Lambda$ case.
 
Pressure as a function baryon mass density is displayed in Figure {\ref{pres}}
for temperatures $T=$ 1, 10 and 100 MeV and proton fractions $Y_p=$ 0.1, 0.3 
and 0.5. Just like the free energy case, we find the hyperon EoS with and 
without $\phi$ mesons at high densities and temperatures to be softer than the nuclear
EoS. Furthermore, the hyperon EoS in the np$\Lambda \phi$ case is stiffer than 
the hyperon EoS in the np$\Lambda$ case. It is worth noting here that
there is no kink or jump in pressure when $\Lambda$s appear in the system. This
indicates that it is a smooth transition from nuclear to hyperon matter.

Figure {\ref{entp}} demonstrates the behavior of entropy per baryon as a
function of baryon mass density. We consider the same values of temperatures
and proton fractions as before. For low temperatures, there is not much 
difference between the results with or without $\Lambda$ hyperons. 
Note that the kinks at low densities originate from changes in the nuclear
composition which are related to nuclear shell effects.
There are some effects of $\Lambda$ hyperons for higher baryon densities at 
$T=$ 10 MeV . As the temperature increases to $T=$ 100 MeV, this difference is 
pronounced. In this case, the entropy per baryon including $\Lambda$
hyperons is higher than that of the nuclear matter. However, we cannot 
differentiate between the results of hyperon matter with and without $\phi$
mesons.  

Examples of data from hyperon EoS tables with and without $\phi$ mesons are 
recorded in
Tables {\ref{table3}} - {\ref{table6}}. For $T=$ 0.1, 10 and 100 MeV, selected 
rows of the main tables with fixed values of $Y_p$ and baryon mass density 
($\rho_B$) are displayed in those tables. The various quantities are 
explained in Appendix \ref{app_entries}}. Two variants of the hyperon EoS tables 
with (BHB$\Lambda \phi$) and without (BHB$\Lambda$) $\phi$ mesons 
in binary as well as Shen98 formats are available 
online.\footnote{See \texttt{http://phys-merger.physik.unibas.ch/\midtilde hempel/eos.html}\label{eospage}.} 
Both hyperon EoS tables are also available in the comprehensive 
CompOSE EoS database\footnote{See \texttt{http://compose.obspm.fr} and \citet{composemanual}.} 
as well as on the stellarcollapse.org 
Web site.\footnote{See \texttt{http://stellarcollapse.org/equationofstate}.}
Tables in Shen98 
format do not include electrons, positrons, and photons whereas binary data
files of EoS tables take into account the contributions of electrons, positrons
and photons. Further details are given on the Web site 
of footnote \ref{eospage}.

\section{Summary and Conclusions}       
We have constructed hyperon EoS tables including $\Lambda$ hyperons for 
supernova simulations and neutron stars in a density 
dependent relativistic mean field model. We also take into account the
$\Lambda$-$\Lambda$ interaction mediated by $\phi$ mesons in this calculation.
The nuclear statistical equilibrium model of \citet{hs1} is adopted for the
description of matter made of light and heavy nuclei coexisting with unbound
nucleons below saturation densities and temperatures up to 50~MeV.
We have denoted the calculation including $\Lambda$ hyperons without $\phi$ 
mesons as the np$\Lambda$ case and that of the $\Lambda$ hyperons with $\phi$ mesons as
the np$\Lambda \phi$ case.  
The DD2 parameter set \citet{typ10} has been used in this calculation
for the nucleons. The vector meson - 
$\Lambda$ hyperon couplings are obtained from the SU(6) symmetry relations of
the quark model, whereas
the scalar meson - $\Lambda$ hyperon coupling is determined from the potential
depth of the $\Lambda$ hyperon in nuclear matter at the saturation density 
of $-30$ MeV which is extracted from the experimental binding energies of 
$\Lambda$ hypernuclei. 
The system is populated with $\Lambda$s using the equilibrium condition
$\mu_n = \mu_{\Lambda}$.
The contribution of $\Lambda$s is considered in our calculation when its
corresponding EoS gives a lower free energy 
than the EoS of only nuclei and nucleons
and when the $\Lambda$ mass 
fraction exceeds $10^{-5}$
at the same time. It is noted that the fraction of $\Lambda$ hyperons
is negligible at low-density and low-temperature domains. The population of 
$\Lambda$ hyperons grows in uniform matter at the cost of neutrons at high 
density. A significant fraction of thermal $\Lambda$ hyperons is populated in
the system at higher temperatures.      

The free energy of the system including $\Lambda$ hyperons is
lower compared to that of the nuclear matter case. However, $\Lambda$ 
hyperon matter involving $\phi$ mesons has higher free energy than that of the
$\Lambda$ hyperon matter without $\phi$ mesons. Regarding the entropy per baryon,
one notices that it is higher than in the case of nuclear matter when more degrees
of freedom in the form of $\Lambda$ hyperons appear in the system. 
This indicates different thermal properties of the EoS, 
which are known to be important for neutron-star mergers \citep{bauswein13,kaplan13}
and black hole formation \citep{hs2}.
We observe 
that the EoS
(pressure versus baryon mass density) of the $\Lambda$ hyperon matter with and 
without $\phi$ mesons is softer than the nuclear EoS. Furthermore, the repulsive
interaction of $\phi$ mesons makes the EoS of the np$\Lambda \phi$ case stiffer 
than that of the np$\Lambda$ case. 
It is important to note that the pressure grows smoothly with baryon density 
even after the appearance of $\Lambda$ hyperons. We did not find any indication 
for a first-order phase transition connected with the appearance of hyperons, 
as discussed, e.g., by \citet{schaffner02,gulmi2012,gulmi2013}.

We have generated two $\Lambda$ hyperon EoS tables with 
(BHB$\Lambda \phi$) and without (BHB$\Lambda$) $\phi$ mesons 
covering temperatures (0.1 -- 158.48 MeV), proton fractions (0.01 -- 0.6), 
and baryon density ($10^{-12}$ -- $\simeq$1 fm$^{-3}$). The EoS tables are 
written in two different formats:
the first format is similar to the one used by \citet{shen}, and the
second one corresponds to
extended tables including electrons, positrons and photons 
in a binary format.
Tables {\ref{table3}} - {\ref{table6}} illustrate 
certain parts of the main tables.

Finally, we impose the charge neutrality and $\beta$-equilibrium in our 
$\Lambda$ hyperon EoS tables and calculate mass-radius relationship of the 
neutron star sequence at $T=0.1$ MeV. We obtain maximum neutron star masses 
2.1 M$_{\odot}$ and 1.95 M$_{\odot}$ corresponding to the $\Lambda$ hyperon EoS
with and without $\phi$ mesons, respectively. The maximum neutron star mass of 
$\Lambda$ hyperon matter including $\phi$ mesons is compatible with the 
recently measured 2.01$\pm$0.04 M$_{\odot}$ neutron star. 

We shall perform supernova simulations with new hyperon EoS tables and publish
those results separately in the future. New hyperon EoS tables will be also 
useful for neutron star merger calculations.

\acknowledgments
Numerical calculations of this work have been partly carried out in the blade 
server of the Astroparticle Physics and Cosmology Division, Saha Institute of 
Nuclear Physics, Kolkata. M.H. acknowledges support from the Swiss National 
Science Foundation and the COST action NewCompstar.

\section*{Appendix}
\appendix 
\section{Description of the Tables in Shen98 Format}
The EoS tables are presented in two different formats
``Extended'' and ``Shen98''
on the Web site \texttt{http://phys-merger.physik.unibas.ch/\midtilde hempel/eos.html}
Here we restrict ourselves to the
description of the latter where the information are stored in a format that 
is similar to the tables of Shen \citep{shen,shen11}, which is 
widely used in many different astrophysical applications.

\subsection{Parameter Grid and Data Structure}
Table \ref{tab:overview} gives an overview of the two hyperonic tables 
BHB$\Lambda$ and BHB$\Lambda \phi$, regarding the constituents considered, and 
the points in the parameter space of temperature, density, and proton fraction 
which were calculated. For density and temperature, we have
a logarithmic spacing, and for the proton fraction it is linear.
Besides the range of density, the two tables cover the same conditions 
(i.e., in temperature and proton fraction).

We arrange the data as follows: We group them in blocks of constant
temperature, starting with the lowest value. Within each temperature block,
we group the data according to the proton fraction, again starting with lowest
values. For given temperature and proton fraction, we list all baryon number 
densities with increasing values.

\subsection{Entries of the Tables}
\label{app_entries}
For each grid point specified by density, temperature, and electron fraction, 
there are 20 
different thermodynamic quantities in the tables. 
Those thermodynamic quantities are explained below.
Note that only baryonic contributions to 
different quantities are recorded. The contributions of photons, electrons, 
positrons, and neutrinos are to be added separately. 

\begin{enumerate}
 
\item Logarithm of baryon mass density (log$_{10} (\rho_B)$ [g/cm$^3$]). 

The baryon mass density is defined as the baryon number density multiplied by 
the value of the atomic mass unit $m_u = $ 931.49432 MeV.

\item Baryon number density ($n_B$ [fm$^{-3}$]).

\item Logarithm of total proton fraction (log$_{10} (Y_p)$ []) .

\item Total proton fraction ($Y_p$ []).

Note that the total proton fraction $Y_p$ is given by all protons 
(i.e., free and bound in nuclei) and thus is equal to the electron fraction
to obtain charge neutrality.

\item Free energy per baryon ($F$ [MeV]).

Free energy per baryon relative to 938 MeV is defined by
\begin{equation}
F = \frac{f}{n_B} - 938~.
\end{equation}
We have chosen the reference value of 938 MeV because it 
was also used in the original table of \citet{shen}. Note that this value
is otherwise completely arbitrary and not used in the EoS calculations.

\item Internal energy per baryon ($E_{\rm int}$ [MeV]).

$E_{\rm int}$ relative to $m_u$ is defined by
\begin{equation}
E_{\rm int} = \frac{\epsilon}{n_B} - m_u~.
\end{equation}

\item Entropy per baryon ($S$ [$k_B$]).

\item Average mass number of heavy nuclei ($\bar A$ []).

This is defined as
$\bar A = \sum_{A,Z \geq 6}A n_{A,Z} / \sum_{A,Z\geq 6} n_{A,Z}$. 

\item Average charge number of heavy nuclei ($\bar Z$ []).

This is defined as
$\bar Z = \sum_{A,Z \geq 6}Z n_{A,Z} / \sum_{A,Z\geq 6} n_{A,Z}$. 

Note that $\bar Z$ and $\bar A$ are set to zero if $X_A=0$, i.e.,
if no heavy nuclei are present.

\item Nucleon effective Dirac mass ($m^*$ [MeV]).

In the RMF calculation, we use separate values for neutron and proton masses. 
However, we store only the average value of neutron and proton effective Dirac masses.

\item Mass fraction of unbound neutrons ($X_n$ []). 

This is defined as $X_n = n_n/n_B$.

\item Mass fraction of unbound protons ($X_p$ []). 

It is given by $X_p = n_p/n_B$.

\item Mass fraction of light nuclei ($X_a$ []).

This is defined as 
$X_a = \sum_{A,Z \leq 5}A n_{A,Z} / n_B$. 

\item Mass fraction of heavy nuclei ($X_A$ []). 

This is defined as 
$X_A = \sum_{A,Z \geq 6}A n_{A,Z} / n_B$. 

\item Baryon pressure ($P$ [MeV/fm$^3$]).

\item Neutron chemical potential relative to neutron rest mass 
($\mu_n-m_n$ [MeV]).

The value of $m_n$ is specified in Table \ref{table2}, which also corresponds
to the value used in our calculations. Note that $\mu_\Lambda = \mu_n$ 
wherever $\Lambda$s are present.

\item Proton chemical potential relative to proton rest mass ($\mu_p-m_p$ [MeV]).

The value of $m_p$ is specified in Table \ref{table2}, which also corresponds
to the value used in our calculations.

\item Average mass number of light nuclei ($\bar a$ []).

This is defined as
$\bar a = \sum_{A,Z \leq 5}A n_{A,Z} / \sum_{A,Z\leq 5} n_{A,Z}$. 

\item Average charge number of light nuclei ($\bar z$ []).

This is defined as
$\bar z = \sum_{A,Z \leq 5}Z n_{A,Z} / \sum_{A,Z\leq 5} n_{A,Z}$. 

Note that $\bar z$ and $\bar a$ are set to zero if $X_a=0$, i.e.,
if no light nuclei are present.

\item Mass fraction of Lambda hyperons ($X_{\Lambda}$ []). 

It is defined as $X_{\Lambda} = n_{\Lambda}/n_B$.
 
\end{enumerate}

\subsection{Accuracy and Consistency of the EoS Tables}

We have performed the following consistency checks on the EoS tables.

\begin{enumerate}

\item Thermodynamic consistency requires 

\begin{equation}
\epsilon = Ts - P + \mu_n (1-Y_p) n_B + \mu_p Y_p n_B~.
\end{equation}

The modulus of the relative thermodynamic accuracy

\begin{equation}
\Delta = \frac {Ts - P + \mu_n (1-Y_p) n_B + \mu_p Y_p n_B} {\epsilon} - 1~,
\end{equation}
is on average $2.0 \times 10^{-6}$ in the EoS tables.

\item  Sum rule of particle fractions is given by

\begin{equation}
X_n + X_p + X_a + X_A + X_{\Lambda} = 1~,
\end{equation}
and is satisfied by the EoS tables 
with an accuracy higher than $1.8\times 10^{-7}$. The average deviation
is $2.6\times 10^{-8}$.

\item We have also checked that the EoS tables fulfill the thermodynamic stability
criteria
\begin{equation}
\frac{\partial { s}}{\partial T} \geq 0 ~,
\end{equation}
and
\begin{equation}
\frac{\partial P^{\rm tot}}{\partial n_B} \geq 0~.
\end{equation}
The second of the two relations is only fulfilled for the total pressure, 
i.e., if the electron contribution is added $P^{\rm tot}=P_e+P$. There are
just ten grid points in the table, where this second relation 
is slightly violated.

\end{enumerate}

Note that all numbers which we have given here are directly calculated 
from the tables in the Shen98 format. Because only seven digits are stored in 
these tables, the highest deviations originate mostly from round-off errors.
In our actual calculations, and in some other binary versions of the tables 
where all quantities are stored with double precision, the accuracy is even 
higher. Note that we also do not apply any smoothing or averaging 
prescription, as is done in, e.g., \citet{horo1,horo2}.

\newpage

\begin{deluxetable}{cccccc} 
\tablecaption{Parameters of Nucleon-Meson Couplings as Defined in Equation (\ref{coef})
and Equation (\ref{rhoc})\label{table1}
}
\tablewidth{0cm}
\tablehead{
\colhead{i} & \colhead{$g_{i N}(n_0)$} & \colhead{$a_{i}$} & \colhead{$b_{i}$} & \colhead{$c_{i}$} & \colhead{$d_{i}$} \\ \hline
}
\startdata
$\sigma$& 10.686681& 1.357630 & 0.634442 & 1.005358 & 0.575810 \\ 
$\omega$& 13.342362& 1.369718 & 0.496475 & 0.817753 & 0.638452 \\ 
$\rho$& 3.626940& 0.518903 & & & \\ 
\enddata
\tablecomments{
The parameters are obtained by reproducing properties of finite nuclei
and the parameter set is known as the DD2 set \citep{typ10}. This parameterization
leads to nuclear saturation properties such as saturation density 
($n_0 = 0.149065$ fm$^{-3}$), binding energy (16.02 MeV), incompressibility of 
matter (242.7 MeV), symmetry energy (31.67 MeV), and its slope (55.03 MeV). 
All parameters are dimensionless. 
}
\end{deluxetable}

\begin{table}
\caption{Masses of Baryons and Mesons in Units of MeV Used in This Calculation
}
\begin{center}
\begin{tabular}{ccccccc} 
\hline\hline
Neutron & Proton & $\Lambda$ & $\sigma$ & $\omega$ & $\rho$ & $\phi$ \\ \hline
939.56536& 938.27203&1115.7& 546.212459 & 783.0 & 763.0 & 1020.0 \\ \hline
\label{table2}
\end{tabular}
\end{center}
\end{table}

\begin{deluxetable}{ccccc}
\tablecaption{Overview of the EoS Tables Presented in This Paper.
\label{tab:overview}}
\tablewidth{0cm}
\tablehead{
\colhead{} & \colhead{} &\colhead{BHB$\Lambda$} & \colhead{BHB$\Lambda\phi$}
}
\startdata
 Constituents & Uniform matter & $n$, $p$, $\Lambda$      & $n$, $p$, $\Lambda$ \\
          & Non-uniform matter & $n$, $p$, $\{A,Z\}$ & $n$, $p$, $\{A,Z\}$ \\
\hline
 $T$ & Range            & $ -1.0 \leq \log_{10}(T) \leq 2.2 $
                        & $ -1.0 \leq \log_{10}(T) \leq 2.2 $ \\
(MeV) & Grid spacing    & $\Delta \log_{10}(T)=0.04$
                        & $\Delta \log_{10}(T)=0.04$ \\
     & Points           & 81 & 81 \\
\hline
 $Y_p$ & Range          & $ 0.01 \leq Y_p \leq 0.60 $
                        & $ 0.01 \leq Y_p \leq 0.60 $  \\
 ()    & Grid spacing   & $\Delta Y_p=0.01$ 
                        & $\Delta Y_p=0.01$ \\
     & Points           & 60 & 60 \\
\hline
 $n_B$ & Range       & $-12 \leq \log_{10}(n_B) \leq 0 $
                        & $-12 \leq \log_{10}(n_B) \leq 0.08 $ \\
 $\rm{(fm^{-3})}$ & Grid spacing & $\Delta \log_{10}(n_B)= 0.04 $
                                & $\Delta \log_{10}(n_B)= 0.04 $ \\
     & Points                   & 301 & 303 \\
\enddata
\tablecomments{
In both tables, non-uniform matter is modeled as a mixture of
free neutrons ($n$), free protons ($p$), and an ensemble of heavy nuclei 
($\{A,Z\}$). Uniform matter consists in general of neutrons, protons, and 
$\Lambda$s ($\Lambda$), whereas $\Lambda$s are only considered, when the conditions 
described in Section~\ref{match} are met. Besides the range of density, 
the two tables cover the same conditions (i.e., in temperature
and proton fraction).}
\end{deluxetable}

\newpage
\begin{deluxetable}{cccccccccc}
\tabletypesize{\tiny}
\tablecaption{
Data from BHB$\Lambda$ EoS Table for $T=0.1$ MeV and 
$Y_p=$ 0.01 and 0.5
\label{table3}}
\tablewidth{0pt}
\tablehead{
 \colhead{$\log_{10}(\rho_B)$} & \colhead{$n_B$} & \colhead{$\log_{10}(Y_p)$} &
 \colhead{$Y_p$} &
 \colhead{$F$}     &  \colhead{$E_{\rm{int}}$}   & \colhead{$S$}   &
 \colhead{$\bar A$}     &  \colhead{$\bar Z$}    & \colhead{$m^{*}$}       
 \\
 \colhead{($\rm{g\,cm^{-3}}$)}     &  \colhead{($\rm{fm^{-3}}$)} & \colhead{}   &
 \colhead{}  &
 \colhead{(MeV)}   &  \colhead{(MeV)}  &  \colhead{($k_B$)}  &
 \colhead{}        &  \colhead{}       &  \colhead{(MeV)}    
}
\startdata
 3.22025 & 1.0E-12 & -2 & 0.01 & -0.385878 & 7.97106 & 18.5126 & 76 &  26 & 939.565 \\ 
14.2203 & 0.1 & -2 & 0.01 & 11.5293 &18.0360 & 0.010149& 0 &  0 & 631.434\\ 
15.2203 & 1.0 & -2 & 0.01 & 288.255 &294.761 & 0.260104E-02& 0 &  0 & 9.95451 \\ \hline 
3.22025 & 1.0E-12 & -0.30103& 0.5 &-7.77206&-1.2157&0.506831& 56 & 28 &938.272 \\ 
14.2203 & 0.1 & -0.30103& 0.5 & -13.6378&-7.13086& 1.29635& 0 &  0 &625.119 \\ 
15.2203 & 1.0 & -0.30103& 0.5 & 305.76  &312.266 & 0.283803E-02& 0 &  0 &52.0946 
\enddata
\tablecomments{This table gives the values of quantities of six single 
rows
in the so-called Shen-98-format, as specified in Appendix \ref{app_entries}. 
The complete hyperon table without $\phi$ mesons 
(BHB$\Lambda$) is available at 
{\texttt{http://phys-merger.physik.unibas.ch/\midtilde hempel/eos/v1.0/bhb\_l\_frdm\_shen98format.zip}}.
Data points with less digits are shown here for guidance regarding its 
form and content.
}
\end{deluxetable}
\begin{deluxetable}{cccccccccc}
\tabletypesize{\tiny}
\tablecaption{Continuation of Table \ref{table3}\label{table3b}}
\tablewidth{0pt}
\tablehead{
 \colhead{$X_n$}   &  \colhead{$X_p$}  & \colhead{$X_{a}$}    & \colhead{$X_A$} &
 \colhead{$P$}     &  \colhead{$\mu_n$}& \colhead{$\mu_p$}         &
 \colhead{$\bar a$}           & \colhead{$\bar z$}   
 & \colhead{$X_{\Lambda}$}
 \\
 \colhead{}        &  \colhead{}       &  \colhead{}         &  \colhead{} &
 \colhead{($\rm{MeV\,fm^{-3}}$)} &  \colhead{(MeV)} &  \colhead{(MeV)} &
 \colhead{}   &  \colhead{}
 &  \colhead{}
}
\startdata
 0.970769 & 0 & 0 & 0.02923 & 0.971141E-13 & -1.65569& -20.2045 & 3.01 & 1.01& 0 \\ 
 0.99 & 0.01 & 0 & 0 & 0.949866& 20.5277& -84.6839 & 0 & 0& 0 \\ 
 0.248928& 0.01 & 0 & 0 & 336.830&626.015& 377.824 & 0 & 0& 0.741072\\ \hline 
 0& 0& 0& 1 & 0.153089E-14&-13.739&-3.63947& 4 & 2& 0\\ 
 0.5& 0.5 & 0 & 0 & -0.609824&-20.6929&-20.6166& 0 & 0& 0.0\\ 
 0.088970& 0.5 & 0 & 0 & 491.997&693.182& 900.493 & 0 & 0& 0.411030
\enddata
\end{deluxetable}
\begin{deluxetable}{cccccccccc}
\tabletypesize{\tiny}
\tablecaption{Same as Table {\ref{table3}}, but for $T=100$ MeV
\label{table4}}
\tablewidth{0pt}
\tablehead{
 \colhead{$\log_{10}(\rho_B)$} & \colhead{$n_B$} & \colhead{$\log_{10}(Y_p)$} &
 \colhead{$Y_p$} &
 \colhead{$F$}     &  \colhead{$E_{\rm{int}}$}   & \colhead{$S$}   &
 \colhead{$\bar A$}     &  \colhead{$\bar Z$}    & \colhead{$m^{*}$}
 \\
 \colhead{($\rm{g\,cm^{-3}}$)}     &  \colhead{($\rm{fm^{-3}}$)} & \colhead{}   &
 \colhead{}  &
 \colhead{(MeV)}   &  \colhead{(MeV)}  &  \colhead{($k_B$)}  &
 \colhead{}        &  \colhead{}       &  \colhead{(MeV)}
}
\startdata
3.22025 & 1.0E-12 & -2 & 0.01 & -2.13589E+10 & 2.40721E+11 & 2.62080E+09& 0 &0& 773.734\\ 
14.2203 & 0.1 & -2 & 0.01 & -279.137 & 206.844 & 4.79475& 0 &0& 542.722 \\ 
15.2203 & 1.0 & -2 & 0.01 & 165.051 &411.954 & 2.40397& 0 &  0 &19.9808 \\ \hline
3.22025 & 1.0E-12 & -0.30103& 0.5 &-2.13589E+10&2.40721E+11&2.62080E+09& 0 &  0 &854.728 \\ 
14.2203 & 0.1 & -0.30103& 0.5 & -341.599 &170.628 & 5.05722& 0 &  0 &619.924 5\\ 
15.2203 & 1.0 & -0.30103& 0.5 & 177.458 &423.538 & 2.39574& 0 &  0 &71.3397 
\enddata
\tablecomments{ The complete hyperon table without $\phi$ mesons 
(BHB$\Lambda$) is available 
at {\texttt{http://phys-merger.physik.unibas.ch/\midtilde hempel/eos/v1.0/bhb\_l\_frdm\_shen98format.zip}}.
Data points with less digits are shown here for guidance regarding its 
form and content.
}
\end{deluxetable}
\begin{deluxetable}{cccccccccc}
\tabletypesize{\tiny}
\tablecaption{Continuation of Table \ref{table4}.\label{table4b}}
\tablewidth{0pt}
\tablehead{
 \colhead{$X_n$}   &  \colhead{$X_p$}  & \colhead{$X_{a}$}    & \colhead{$X_A$} &
 \colhead{$P$}     &  \colhead{$\mu_n$}& \colhead{$\mu_p$}         &
 \colhead{$\bar a$}           & \colhead{$\bar z$}   
 & \colhead{$X_{\Lambda}$}
 \\
 \colhead{}        &  \colhead{}       &  \colhead{}         &  \colhead{} &
 \colhead{($\rm{MeV\,fm^{-3}}$)} &  \colhead{(MeV)} &  \colhead{(MeV)} &
 \colhead{}   &  \colhead{}
 &  \colhead{}
}
\startdata
0.814211 & 0.01 & 0 & 0 & 0.0213589 & -939.565& -938.272& 0 & 0& 0.175788\\ 
0.772041 & 0.01 & 0 & 0 & 11.4656 & -161.377& -627.033& 0 & 0& 0.217959\\ 
0.274440& 0.01 & 0 & 0 & 407.159&573.612& 278.197 & 0 & 0& 0.715560\\ \hline
0.411164& 0.5 & 0 & 0 & 0.0213589&-939.565&-938.272 & 0 & 0& 0.0888365\\ 
0.408025& 0.5 & 0 & 0 & 11.2739&-242.685&-216.873 & 0 & 0& 0.091975\\ 
0.105272& 0.5 & 0 & 0 & 554.206&614.827& 846.663 & 0 & 0& 0.394728
\enddata
\end{deluxetable}
\begin{deluxetable}{cccccccccccccccccccc}
\tabletypesize{\tiny}
\tablecaption{
Same as Table {\ref{table3}}, but now for the BHB$\Lambda \phi$ EoS table for 
T = 10 MeV.
\label{table5}}
\tablewidth{0pt}
\tablehead{
 \colhead{$\log_{10}(\rho_B)$} & \colhead{$n_B$} & \colhead{$\log_{10}(Y_p)$} &
 \colhead{$Y_p$} &
 \colhead{$F$}     &  \colhead{$E_{\rm{int}}$}   & \colhead{$S$}   &
 \colhead{$\bar A$}     &  \colhead{$\bar Z$}    & \colhead{$m^{*}$} 
 \\
 \colhead{($\rm{g\,cm^{-3}}$)}     &  \colhead{($\rm{fm^{-3}}$)} & \colhead{}   &
 \colhead{}  &
 \colhead{(MeV)}   &  \colhead{(MeV)}  &  \colhead{($k_B$)}  &
 \colhead{}        &  \colhead{}       &  \colhead{(MeV)} 
}
\startdata
 3.22025 & 1.0E-12 & -2 & 0.01 & -243.554 &23.2555 & 26.0303 & 8 &  6 & 939.552\\ 
14.2203 & 0.1 & -2 & 0.01 &  7.01729&22.2363 & 0.871332& 19.81 &  6& 633.164\\ 
15.2203 & 1 & -2 & 0.01 & 331.522 &340.708 & 0.268107& 0 &  0 &29.9543\\ \hline 
3.22025 & 1.0E-12 & -0.30103& 0.5 &-250.549&22.6219&26.6665& 8 & 6 &938.919\\ 
14.2203 & 0.1 & -0.30103& 0.5 & -19.8927&-1.34810& 1.20389& 0 &  0 &627.659\\ 
15.2203 & 1.0 & -0.30103& 0.5 & 319.375 &328.752 & 0.287131& 0 &  0 &61.3811
\enddata
\tablecomments{The complete hyperon table with $\phi$ mesons 
(BHB$\Lambda \phi$) is available at
{\texttt{http://phys-merger.physik.unibas.ch/\midtilde hempel/eos/v1.0/bhb\_lp\_frdm\_shen98format.zip}}.
Data points with less digits are shown here for guidance regarding its 
form and content.
}
\end{deluxetable}
\begin{deluxetable}{cccccccccccccccccccc}
\tabletypesize{\tiny}
\tablecaption{Continuation of Table \ref{table5}.\label{table5b}}
\tablewidth{0pt}
\tablehead{
 \colhead{$X_n$}   &  \colhead{$X_p$}  & \colhead{$X_{a}$}    & \colhead{$X_A$} &
 \colhead{$P$}     &  \colhead{$\mu_n$}& \colhead{$\mu_p$}         &
 \colhead{$\bar a$}           & \colhead{$\bar z$}   
 & \colhead{$X_{\Lambda}$}
 \\
 \colhead{}        &  \colhead{}       &  \colhead{}         &  \colhead{} &
 \colhead{($\rm{MeV\,fm^{-3}}$)} &  \colhead{(MeV)} &  \colhead{(MeV)} &
 \colhead{}   &  \colhead{}
 &  \colhead{}
}
\startdata
 0.99 & 0.01 & 0 & 0 & 1.0E-11 & -234.647& -280.577 & 2 & 1& 0 \\ 
 0.989845 & 0.009922 & 0.000232 & 0 & 1.32322& 19.9713& -107.445 & 3.11& 1.04& 0 \\ 
 0.414247& 0.01 & 0 & 0 & 448.471&781.568& 468.839 & 0 & 0& 0.575753\\ \hline 
 0.5& 0.5& 0& 0 & 1.0E-11&-241.478&-241.457& 2 & 1& 0\\ 
 0.5& 0.5 & 0 & 0 & -0.069936&-21.5511&-21.4705& 0 & 0& 0.0\\ 
 0.149693& 0.5 & 0 & 0 & 533.660&773.293& 930.939 & 0 & 0& 0.350307
\enddata
\end{deluxetable}
\begin{deluxetable}{cccccccccccccccccccc}
\tabletypesize{\tiny}
\tablecaption{Same as Table {\ref{table5}}, but for $T=100$ MeV.
\label{table6a}}
\tablewidth{0pt}
\tablehead{
 \colhead{$\log_{10}(\rho_B)$} & \colhead{$n_B$} & \colhead{$\log_{10}(Y_p)$} &
 \colhead{$Y_p$} &
 \colhead{$F$}     &  \colhead{$E_{\rm{int}}$}   & \colhead{$S$}   &
 \colhead{$\bar A$}     &  \colhead{$\bar Z$}    & \colhead{$m^{*}$}
 \\
 \colhead{($\rm{g\,cm^{-3}}$)}     &  \colhead{($\rm{fm^{-3}}$)} & \colhead{}   &
 \colhead{}  &
 \colhead{(MeV)}   &  \colhead{(MeV)}  &  \colhead{($k_B$)}  &
 \colhead{}        &  \colhead{}       &  \colhead{(MeV)}
}
\startdata
 3.22025 & 1.0E-12 & -2 & 0.01 & -2.13590E+10 & 2.40721E+11 & 2.62080E+09& 0 &0& 773.746 \\ 
14.2203 & 0.1 & -2 & 0.01 & -278.401 & 205.785 & 4.77680& 0 &0& 549.326 \\ 
15.2203 & 1.0 & -2 & 0.01 & 208.007 &455.691 & 2.41179& 0 &  0 &39.9208 \\ \hline
3.22025 & 1.0E-12 & -0.30103& 0.5 &-2.13590E+10&2.40721E+11&2.62080E+09& 0 &  0 &854.734 \\ 
14.2203 & 0.1 & -0.30103& 0.5 & -341.465 &170.346 & 5.05305& 0 &  0 &621.205 \\ 
15.2203 & 1.0 & -0.30103& 0.5 & 191.740 &438.178 & 2.39932& 0 &  0 &79.6312 
\enddata
\tablecomments{ The complete hyperon table with $\phi$ mesons 
(BHB$\Lambda \phi$) is available at
{\texttt{http://phys-merger.physik.unibas.ch/\midtilde hempel/eos/v1.0/bhb\_lp\_frdm\_shen98format.zip}}.
Data points with less digits are shown here for guidance regarding its 
form and content.
}
\end{deluxetable}
\begin{deluxetable}{cccccccccccccccccccc}
\tabletypesize{\tiny}
\tablecaption{Continuation of Table \ref{table6a}.\label{table6}}
\tablewidth{0pt}
\tablehead{
 \colhead{$X_n$}   &  \colhead{$X_p$}  & \colhead{$X_{a}$}    & \colhead{$X_A$} &
 \colhead{$P$}     &  \colhead{$\mu_n$}& \colhead{$\mu_p$}         &
 \colhead{$\bar a$}           & \colhead{$\bar z$}   
 & \colhead{$X_{\Lambda}$}
 \\
 \colhead{}        &  \colhead{}       &  \colhead{}         &  \colhead{} &
 \colhead{($\rm{MeV\,fm^{-3}}$)} &  \colhead{(MeV)} &  \colhead{(MeV)} &
 \colhead{}   &  \colhead{}
 &  \colhead{}
}
\startdata
 0.814225 & 0.01 & 0 & 0 & 0.0213589 & -939.565& -938.272& 0 & 0& 0.175775\\ 
 0.782574 & 0.01 & 0 & 0 & 11.5394 & -159.885& -627.353& 0 & 0& 0.207427\\ 
 0.414758& 0.01 & 0 & 0 & 504.552&714.667& 348.584 & 0 & 0& 0.575242\\ \hline
 0.411170& 0.5 & 0 & 0 & 0.0213589&-939.565&-938.272 & 0 & 0& 0.0888298\\ 
 0.410135& 0.5 & 0 & 0 & 11.2864&-242.108&-216.932 & 0 & 0& 0.089865\\ 
 0.153005& 0.5 & 0 & 0 & 588.225&687.012& 871.080 & 0 & 0& 0.346994
\enddata
\end{deluxetable}
\clearpage

\clearpage

\clearpage

\clearpage

\begin{figure}
\epsscale{1.0}
\plotone{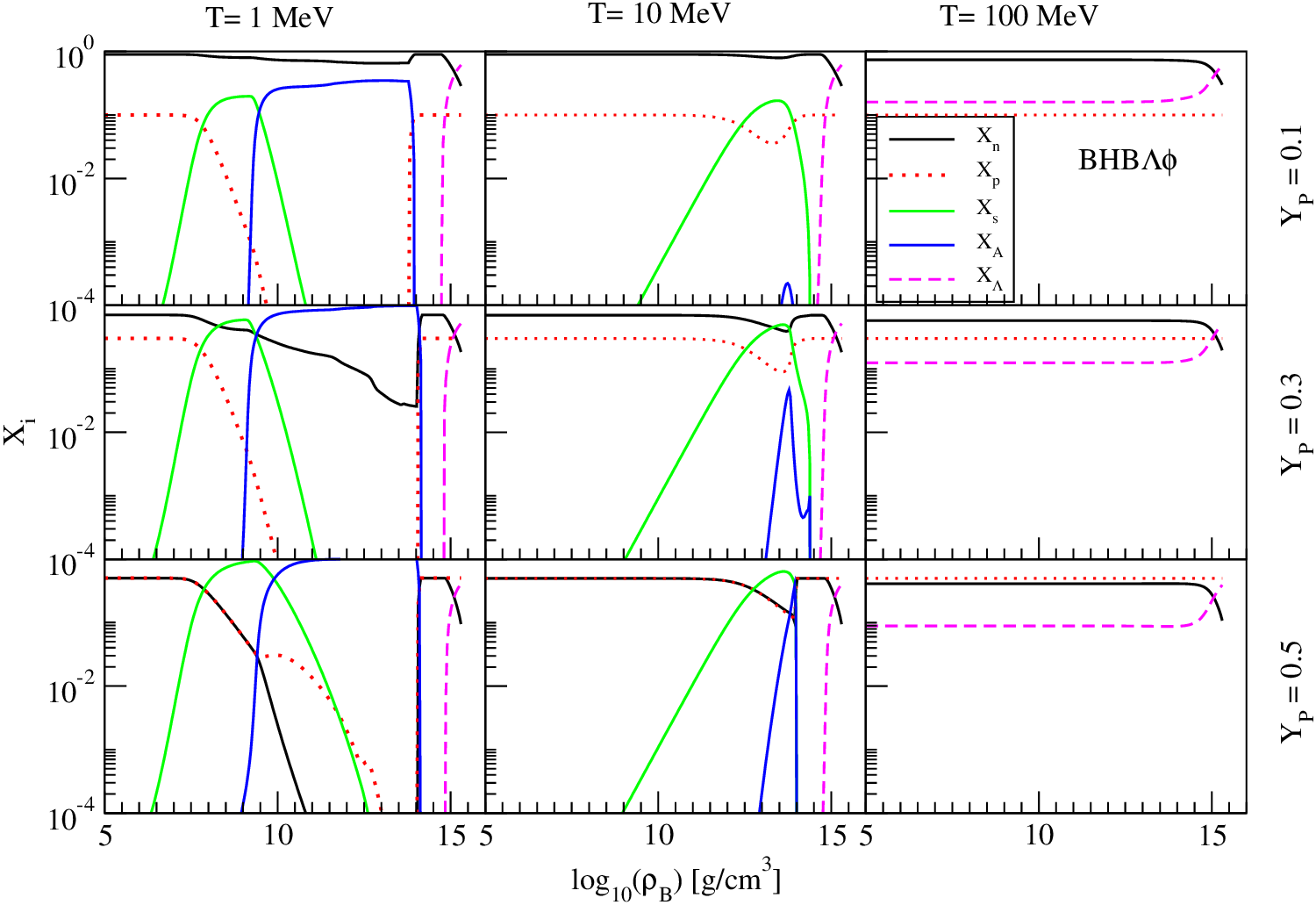}
\caption{ Mass fractions of neutrons ($X_n$), protons ($X_p$), light nuclei 
($X_a$), heavy nuclei ($X_A$), and $\Lambda$ hyperons ($X_{\Lambda}$) are 
plotted as a function of baryon mass density for BHB$\Lambda \phi$
corresponding to the np$\Lambda \phi$ case.
} 
\label{fracp}
\end{figure}

\clearpage

\clearpage

\clearpage

\clearpage

\begin{thebibliography}{}
\bibitem[Antoniadis et al.\ (2013)]{anto} Antoniadis, J., Freire, P. C. C., Wex, N., et al.\
2013, Science, 340, 448
\bibitem[Aparicio (1998)]{aparicio98} Aparicio, J.~M. 1998, 
Astrophys. J. Suppl. Ser. 117, 627
\bibitem[Audi et al.\ (2003)]{audi} Audi, G., Wapstra, A. H., \& Thibault, C. 
2003, Nucl. Phys. A, 729, 337
\bibitem[Banik (2014)]{sb} Banik, S. 2014, Phys. Rev. C, 89, 035807  
\bibitem[Banik \& Bandyopadhyay(2002)]{bani02} Banik, S., \& 
Bandyopadhyay, D. 2002, Phys. Rev. C, 66, 065801
\bibitem[Bauswein et al.\ (2013)]{bauswein13} 
Bauswein, A., Goriely, S., \& Janka, H.-T. 2013, Astrophys. J., 773, 78
\bibitem[Bethe (1990)]{Bet} Bethe, H. A. 1990, Rev. Mod. Phys., 62, 801
\bibitem[Bethe \& Wilson(1985)]{wil} Bethe, H. A., \& Wilson, J. R. 1985, 
Astrophys. J., 295, 14
\bibitem[Blinnikov et al.\ (2011)]{bli} Blinnikov, S. I., Panov, I. V., 
Rudzsky, M. A., \& Sumiyoshi, K. 2011, Astron. Astrophys. 535, A37 
\bibitem[Botvina \& Mishustin (2004)]{botvina04} Botvina, A. S., \& Mishustin, I. N. 2004, Phys. L. B 584, 233
\bibitem[Botvina \& Mishustin (2010)]{botvina10} Botvina, A. S., \& Mishustin, I. N. 2010, Nucl. Phys. A 843, 98
\bibitem[Brat et al.\ (1999)]{brat} Brat, S., Chrien, R. E., Franklin, W. A., et al.\
1999, Phys. Rev. Lett., 83, 5238
\bibitem[Burrows \& Lattimer (1984)]{burrows84} Burrows, A., \& Lattimer, J. M. 1984, Astrophys. J., 285,  294
\bibitem[Buyukcizmeci et al.\ (2013b)]{buyu} Buyukcizmeci, N., Botvina, A. S., 
\& Mishustin, I. N. 2014, Astrophys. J., 789, 33
\bibitem[Buyukcizmeci et al.\ (2013a)]{buyu_compare} 
Buyukcizmeci, N., Botvina, A. S., Mishustin, I. N., et al.\, 2013, Nucl. Phys. A, 907, 13.
\bibitem[Colucci \& Sedrakian (2013)]{colu} Colucci, G., \& Sedrakian, A.
2013, Phys. Rev. C, 87, 055806
\bibitem[Constantinou et al.\ (2014)]{const} Constantinou, C., Muccioli, B., 
Prakash, M., \& Lattimer, J. M. 2014, Phys. Rev. C, 89, 065802
\bibitem[Dover \& Gal (1985)]{dov} Dover, C. B.,
\& Gal, A. 1985, Prog. Part. Nucl. Phys., 12, 171 
\bibitem[F\'{a}i \& Randrup (1982)]{fai82} F\'{a}i, G., \&  Randrup, J. 1982 Nucl. Phys. A, 381, 557
\bibitem[Fischer et al.\ (2014)]{fis2} Fischer, T., Hempel, M., Sagert, I., 
Suwa, Y., \& Schaffner-Bielich, J. 2014, Eur. Phys. J. A, 50, 46 
\bibitem[Fischer et al.\ (2011)]{fis1} Fischer, T., Sagert, I., Pagliara, G., et al.\
2011, Astrophys. J. Suppl. Ser., 194, 39
\bibitem[Gal \& Millener (2011)]{gal} Gal, A., 
\& Millener, D. 2011, Phys. Lett. B, 701, 342 
\bibitem[Glendenning (1985)]{glendenning85} Glendenning, N. K. 1985, 
Astrophys. J. 293, 470.
\bibitem[Glendenning (2000)]{glen} Glendenning, N. K. 
2000, Compact Stars (New York: Springer) 
\bibitem[Gong et~al. (2001)]{gong01} Gong, Z., Zejda, L., D{\"a}ppen, W., \&
Aparicio, J.~M. 2001, Computer Physics Communications 136, 294 
\bibitem[Gulminelli (2012)]{gulmi2012} Gulminelli, F., Raduta, Ad. R., \& Oertel, M.
2012, Phys. Rev. C, 86, 025805
\bibitem[Gulminelli (2013)]{gulmi2013} Gulminelli, F., Raduta, Ad. R., Oertel, M., \& Margueron, J.
2013, Phys. Rev. C, 87, 055809
\bibitem[Gusakov et al.\ (2014)]{gus} Gusakov, M. E., Haensel, P., 
\& Kantor E. M. 2014, Mon. Not. R. Astron. Soc., 439, 318
\bibitem[Hempel et al.\ (2012)]{hs2} Hempel, M., Fischer, T., 
Schaffner-Bielich, J., \& Liebend\"orfer, L. 2012, Astrophys. J., 748, 70 
\bibitem[Hempel \& Schaffner-Bielich (2010)]{hs1} Hempel, M., 
\& Schaffner-Bielich, J. 2010, Nucl. Phys. A, 837, 210 
\bibitem[Hempel et al.\ (2011)]{hemp} Hempel, M., Schaffner-Bielich, J., 
Typel, S., \& R\"opke, G. 2011, Phys. Rev. C, 84, 055804
\bibitem[Hofmann et al.\ (2001a)]{hof1} Hofmann, F., Keil, C. M., 
\& Lenske, H. 2001a, Phys. Rev. C, 64, 034314
\bibitem[Hofmann et al.\ (2001b)]{hof2} Hofmann, F., Keil, C. M., 
\& Lenske, H. 2001b, Phys. Rev. C, 64, 025804
\bibitem[Ishizuka et al.\ (2008)]{ishi} Ishizuka, C., Ohnishi, A., 
Tsubakihara, K., Sumiyoshi, K., \& Yamada, S. 2008, J. Phys. G, 35, 085201 
\bibitem[Janka (2012)]{janka12} Janka, H.-T. 2012, Annu. Rev. Nucl. Part. Sci., 62, 407
\bibitem[Juodagalvis et al.\ (2010)]{juodagalvis10} Juodagalvis, A., Langanke, K., Hix, W. R., Martinez-Pinedo, G. \&  Sampaio, J. M. 
2010, Nucl. Phys. A, 848, 454
\bibitem[Kaplan et al.\ (2013)]{kaplan13}
Kaplan, J. D., Ott, C. D., O'Connor, E. P., et al.\ 2014, Astrophys. J., 790, 19
\bibitem[Lastowiecki et al.\ (2012)]{last} Lastowiecki, R., Blaschke, H., 
Grigorian, H., \& Typel, S. 2012, Acta Phys. Polon. Suppl., 5, 535
\bibitem[Lattimer \& Lim (2013)]{jim} Lattimer, J. M., \& Lim, Y. 2013, 
Astrophys. J., 771, 51 
\bibitem[Lattimer \& Swesty (1991)]{ls} Lattimer, J. M., \& Swesty, F. D. 1991, 
Nucl. Phys. A, 535, 331 
\bibitem[Lopes \& Menezes (2013)]{lopes} Lopes, L. L., \& Menezes, D. P. 2014, Phys. Rev. C, 89, 02805
\bibitem[Mares et al.\ (1995)]{mar} Mares, J., 
Friedman, E., Gal, A., \& Jennings, B. 1995, Nucl. Phys. A, 594, 311
\bibitem[Millener et al.\ (1988)]{mil} Millener, D. J., Dover, C. B.,
\& Gal, A. 1988, Phys. Rev. C, 38, 2700
\bibitem[M\"oller et al.\ (1995)]{moel} M\"oller, P., Nix, J. R., Myers, W. D., 
\& Swiatecki, W. J. 1995, Atomic Data and Nuclear Data Tables, 59, 185
\bibitem[Nakazawa et al.\ (2010)]{nakaza} Nakazawa, K. KEK-E176 collaborators, E373 collaborators, \& J-PARC E07 collaborators,  
2010, Nucl. Phys. A, 835, 207
\bibitem[Nakazato et al.\ (2012)]{naka} Nakazato, K., Furusawa, S., 
Sumiyoshi, K., et al.\ 2012 Astrophys. J., 745, 197 
\bibitem[Nakazato et al.\ (2008)]{naka08} Nakazato, K., 
Sumiyoshi, K., \& Yamada, S. 2008, Phys. Rev. D, 77, 103006
\bibitem[Nordhaus et al.\ (2010)]{nord} Nordhaus, J., Burrows, A., Almgren, A.,
\& Bell, J. 2010, Astrophys. J., 720, 694
\bibitem[Oertel et al.\ (2012)]{oertel12} Oertel, M., Fantina, A. F., 
\& Novak, J. 2012, Phys. Rev. C, 85, 055806
\bibitem[Pagliara et al.\ (2014)]{pagliara14} Pagliara, G., Drago, A., Lavagno, A., \& Pigato, D. 2014, Acta Phys. Polon. Supp., 7, 451
\bibitem[Peres et al.\ (2013)]{peres} Peres, B., Oertel, M., \& Novak, J.
2013, Phys. Rev. D, 87, 043006
\bibitem[Provid\^encia (2013)]{provi2013}Provid\^encia, C., \& Rabhi, A., 
2013, Phys. Rev. C 87, 055801
\bibitem[Qin et al.\ (2012)]{qin12} Qin, L., Hagel, K., Wada, R., et al.\ 2012, Phys. Rev. Lett., 108, 172701
\bibitem[Raduta \& Gulminelli (2010)]{raduta10} Raduta, Ad. R., \& Gulminelli, F.
2010, Phys. Rev. C, 82, 065801
\bibitem[Sagert et al.\ (2009)]{irina} Sagert, I., Fischer, T., Hempel, 
M., Pagliara, G., \& Schaffner-Bielich, J. 2009, Phys. Rev. Lett., 102, 081101
\bibitem[Schaffner \& Gal (2000)]{sch00} 
Schaffner, J., \& Gal, A. 2000, Phys. Rev. C, 62, 034311
\bibitem[Schaffner \& Mishustin (1996)]{sch} 
Schaffner, J., \& Mishustin, I. N. 1996, Phys. Rev. C, 53, 1416
\bibitem[Schaffner et al.\ (1992)]{sch92} 
Schaffner, J., St\"ocker, H., \& Greiner, C. 1992, Phys. Rev. C, 46, 322
\bibitem[Schaffner-Bielich et al.\ (2002)]{schaffner02}
Schaffner-Bielich, J., Hanauske, M., Stöcker, H., \& Greiner, W. 2002, Phys. Rev. Lett., 89, 171101 
\bibitem[Shen et al.\ (2011b)]{horo2} Shen, G., Horowitz, C. J., 
\& O'Connor, E. 2011a, Phys. Rev. C, 83, 065808
\bibitem[Shen et al.\ (2010)]{horo} Shen, G., Horowitz, C. J., 
\& Teige, S. 2010, Phys. Rev. C, 82, 045802
\bibitem[Shen et al.\ (2011a)]{horo1} Shen, G., Horowitz, C. J., 
\& Teige, S. 2011b, Phys. Rev. C, 83, 035802
\bibitem[Shen et al.\ (1998)]{shen} Shen, H., Toki, H., Oyamatsu,
K., \& Sumiyoshi, K. 1998, Nucl. Phys. A, 637, 435
\bibitem[Shen et al.\ (2011c)]{shen11} Shen, H., Toki, H., Oyamatsu,
K., \& Sumiyoshi, K. 2011c, Astrophys. J. Suppl. Ser., 197, 20
\bibitem[Steiner et al.\ (2013)]{stei} Steiner, A., Hempel, M., 
\& Fischer, T. 2013, Astrophys. J., 774, 17
\bibitem[Sumiyoshi et al.\ (2009)]{sumi} Sumiyoshi, K., Ishizuka, C., 
Ohnishi, A., Yamada, S., \& Suzuki, H. 2009, Astrophys. J., 690, L43
\bibitem[Sumiyoshi \& R\"opke (2008)]{sumi08} Sumiyoshi, K., \& R\"opke, G. 2008, Phys. Rev. C, 77, 055804 
\bibitem[Takahashi et al.\ (2001)]{taka} Takahashi H., Ahn J. K., Akikawa H., 
et al.\ 2001, Phys. Rev. Lett., 87, 212502
\bibitem[Togashi et al.\ (2014)]{toga} Togashi, H., Takano, H., 
Sumiyoshi, K., \& Nakazato, K. 2014, Prog. Theor. Exp. Phys., 023D05
\bibitem[Typel et al.\ (2013)]{composemanual} Typel, S., Oertel, M., Klaehn, T. 2013, arXiv:1307.5715
\bibitem[Typel et al.\ (2010)]{typ10} Typel, S., R\"opke, G., Kl\"ahn, 
T., Blaschke, D., \& Wolter, H. H. 2010, Phys. Rev. C, 81, 015803
\bibitem[Typel \& Wolter (1999)]{typ99} Typel, S., \& Wolter, H. H. 1999, Nucl. 
Phys. A, 656, 331
\bibitem[van Dalen et al.\ (2014)]{vandalen14} van Dalen, E. N. E., Colucci, G., \& Sedrakian, A. 2014, Phys. L. B, 734, 383
\bibitem[Weissenborn et al.\ (2012a)]{weis1} Weissenborn, S., Chatterjee, D., 
\& Schaffner-Bielich, J. 2012a, Nucl. Phys. A, 881, 62
\bibitem[Weissenborn et al.\ (2012b)]{weis2} Weissenborn, S., Chatterjee, D., 
\& Schaffner-Bielich, J. 2012b, Phys. Rev. C, 85, 065802
\bibitem[Wolff \& Hillebrandt (1985)]{wolf} Wolff, R. G. \& Hillebrandt, W.
1985, in Nucleosynthesis - Challenges and New Developments, ed. W. D. Arnett \& 
J. M. Truran (Chicago, IL: Univ. of Chicago), 131
\bibitem[Zimanyi \& Moszkowski (1990)]{zimanyi90} Zimanyi, J., \& Moszkowski, S. A. 1990, Phys. Rev. C, 42, 1416
\end{thebibliography}
\end{document}